\documentclass[
 reprint,
 amsmath,
 amssymb,
 aps,
 superscriptaddress,
]{revtex4-2}

\usepackage[utf8]{inputenc}
\usepackage{graphicx}
\usepackage{hyperref}
\usepackage{gensymb}
\usepackage{siunitx}
\usepackage{threeparttable, tablefootnote}
\usepackage{booktabs}
\usepackage{csquotes}
\usepackage[export]{adjustbox}
\usepackage{placeins}
\usepackage{setspace}
\usepackage{enumitem}
\usepackage{verbatim}
\usepackage{mathtools}
\usepackage{bbold}
\usepackage[thinc]{esdiff}
\usepackage{amsmath}
\newcommand{\bb}[1]{\mathbf{#1}}
\newcommand{\w}{\omega}
\newcommand{\rabi}{\Omega}
\newcommand{\h}{\hbar}
\newcommand{\kb}[2]{\ket{#1}\bra{#2}}
\usepackage{cleveref}
\usepackage{multirow}

\usepackage[caption=false]{subfig}
\usepackage[usenames,dvipsnames]{color}
\usepackage{braket}

\usepackage{dcolumn}
\usepackage{bm}

\usepackage[utf8]{inputenc}
\usepackage[T1]{fontenc}
\usepackage{mathptmx}
\usepackage{etoolbox}

\newcommand{\ion}[2]{\ensuremath{^{#2}\mathrm{#1}^+}}

\newcommand{\dfh}[0]{5$D_{5/2}$}
\newcommand{\baonethirtyseven}{$^{137}$Ba$^{+}$}

\newcommand{\MITAffiliation}[0]{Center for Ultracold Atoms, Research Laboratory of Electronics, Massachusetts Institute of Technology, Cambridge, Massachusetts 02139, USA}
\newcommand{\LLAffiliation}[0]{Lincoln Laboratory, Massachusetts Institute of Technology, Lexington, Massachusetts 02421, USA}

\begin{document}

\preprint{APS/123-QED}

\title{Efficient Implementation of a Quantum Algorithm with a Trapped Ion Qudit}

\author{Xiaoyang Shi}
\email{shix@mit.edu}
\affiliation{\MITAffiliation}

\author{Jasmine Sinanan-Singh}
\affiliation{\MITAffiliation}

\author{Timothy J. Burke}
\affiliation{\MITAffiliation}

\author{John Chiaverini}
\affiliation{\MITAffiliation}
\affiliation{\LLAffiliation}

\author{Isaac L. Chuang}
\affiliation{\MITAffiliation}


%

\begin{abstract}

Demonstration of quantum advantage remains challenging due to the increased overhead of controlling large quantum systems. While significant effort has been devoted to qubit-based devices, qudits ($d$-level systems) offer potential advantages in both hardware efficiency and algorithmic performance. In this paper, we demonstrate multi-tone control of a single trapped ion qudit of up to eight levels, as well as the first implementation of Grover's search algorithm on a qudit with dimension five and eight, achieving operation fidelity of 96.8(3)$\%$ and 69(6)$\%$, respectively, which correspond to 99.9(1)\% and 97.1(3) \% squared statistical overlap (SSO), respectively, with the expected result for a single iteration of the Grover search algorithm. The performance is competitive when compared to qubit-based systems; moreover, the sequence requires only $\mathcal{O}(d)$ single qudit gates and no entangling gates. This work highlights the potential of using qudits for efficient implementations of quantum algorithms.

\end{abstract}
\maketitle








Due to the increased difficulty in manipulating large arrays of individually addressable quantum systems, experimental efforts in quantum computing are limited in qubit number, typically at the scale of hundreds to thousands~\cite{Bluvstein2024, Manetsch2024}.The challenge of scaling quantum systems is twofold. First, increasing the number of physical particles that encode quantum information while maintaining high-fidelity operations is complicated by control limitations~\cite{Bruzewicz2019, Henriet2020} or increased crosstalk~\cite{Krantz2019}. Second, multi-qubit entangling gates required to perform quantum algorithms, such as Grover's search or Shor's factoring algorithm~\cite{Monz2016, Figgatt2017}, are expensive to implement on qubit-based systems. The $n$-qubit entangling gates typically need to be decomposed into $\mathcal{O}(n^2)$ two-qubit gates, when not using ancilla qubits, or  into $\mathcal{O}(n)$ entangling gates with $\mathcal{O}(n)$ ancilla qubits~\cite{Adriano1995,Nielsen_Chuang_2010, Figgatt2017, AbuGhanem2025, Nikolaeva2023, Saeedi2013}. It is therefore challenging to show an advantage over these algorithms' classical counterparts, e.g. when applied to large datasets~\cite{Lubinski2023}.

Qudits with $d>2$ levels could complement the effort to scale quantum information processors, providing new opportunities for hardware efficiency. This is due to their larger Hilbert space when compared to qubits, allowing for more efficient encoding of information using smaller physical systems~\cite{Ashok2000, Ivanov2012}. Additionally, multi-particle entangling gates can be performed more straightforwardly, without the need for ancilla particles~\cite{Amit2022, Nikolaeva2023, Kiktenko2020}. Recent theoretical and experimental demonstrations of the encoding of a logical qubit within a single qudit also show the potential for hardware efficiency in quantum error correction~\cite{Debry2025, Li2025, Chiesa2020}. 


Several physical platforms can support qudit encoding, including neutral atoms (\cite{Ma2023, Omanakuttan2023, Omanakuttan2024}, high-spin nuclei \cite{Yu2025, morello24, Godfrin2017}, superconducting circuits \cite{champion24, Goss2022, Laurin2023, Neeley2009}, photonics \cite{Chi2022, Kues2017} and trapped ions \cite{Aksenov2023, Hrmo2023, Ringbauer2022}. For efficient control of qudits, it is not sufficient to simply adapt qubit-like, pairwise operations---namely, the Given's rotations~\cite{Gavin2005}---as this control scheme suffers from a $\mathcal{O}(d^2)$ scaling in the number of pulses needed to implement an arbitrary unitary~\cite{Ringbauer2022}.
Recent demonstrations in transmon and solid-state qudits~\cite{Neeley2009, champion24, Yu2025} have shown efficient control using a multi-tone drive on qudits with up to 8 levels, requiring only $\mathcal{O}(d)$ pulses to implement an arbitrary unitary operation. Randomized benchmarking and the creation of spin-cat states have also been demonstrated using this multi-tone control scheme, suggesting that it may be possible to realize hardware-efficient, high-fidelity implementations of quantum algorithms using this technique. 

A pioneering implementation of Grover's search with qudits used a similar multi-tone control scheme for a nuclear spin ($I = \frac{3}{2}$) within a Tb$^{3+}$ ion~\cite{Godfrin2017}. In this case, however, due to the lack of a pulse sequence capable of generating an equal superposition of four states with equal phases, the algorithm was implemented only on a $d = 3$ subspace and an algorithm success probability of $\sim80\%$ was achieved, highlighting the challenge of scaling beyond $d = 3$.

\begin{figure}[htp]
    \centering
    \includegraphics[width = 0.45\textwidth]{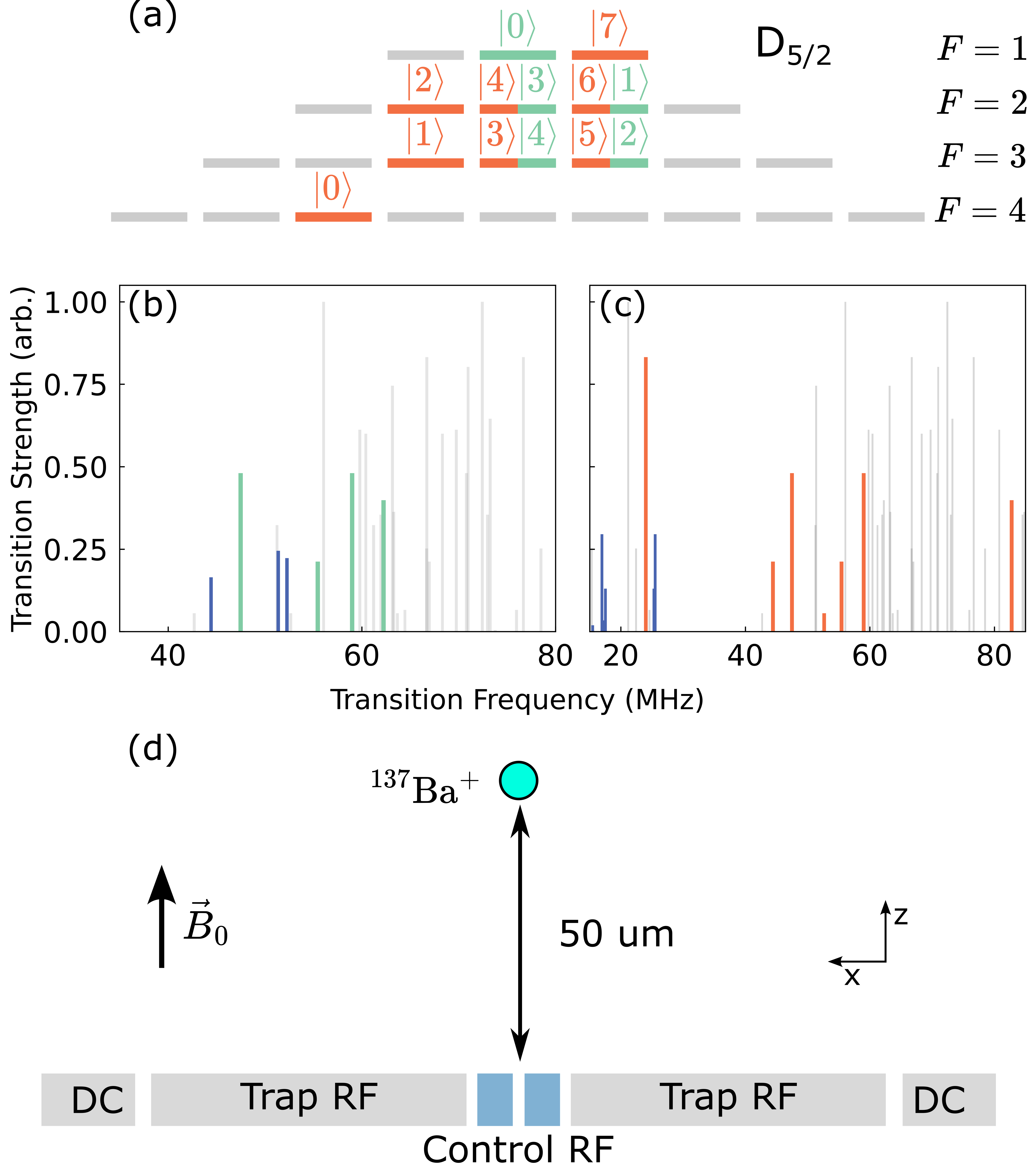}
    \caption{Qudit encoding and experimental setup.  (a) The selected hyperfine states for the encoding of qudits up to either $d = 5$ (green) or $d= 8$ (orange). The transition strength versus transition frequency for these two cases is plotted in (b) and (c). The blue lines represent transitions that involve some qudit states that are off-resonantly driven with a detuning smaller than 10 MHz when all tones are switched on. The gray lines represent all other transitions between the hyperfine levels. (d) The ion trap electrode configuration. The radio-frequency (RF) electrodes used for multi-tone control are indicated in blue and the ion is trapped 50~$\mu$m above the trap surface. The quantization axis is indicated by the direction of the external static magnetic field $\vec{B_{0}}$.}
    \label{fig:levels}
\end{figure}

A promising approach for addressing this  challenge is the use of trapped ions, especially due to the record-high fidelity single- and two-qubit gates achievable in this system~\cite{Loschnauer2024, Ballance2016}. The rich atomic structure also allows for the encoding of high-dimensional qudits~\cite{Low2025}. At the same time, to efficiently implement a quantum algorithm with high fidelity using multi-tone control of trapped-ion qudit systems, a suitable set of states and operations must be determined.  
The control parameters need to be well calibrated, and unitary operations that map the desired algorithm to the qudit manifold must be determined.  

In this paper, we employ the metastable state of \ion{Ba}{137},  a well-established platform for encoding qudits where high-fidelity state preparation and measurement of up to thirteen levels have been demonstrated \cite{Low2025}. We demonstrate the first multi-tone control of a single atomic qudit, utilizing up to eight levels within a \ion{Ba}{137} ion, to realize Grover's search algorithm for $d=5$ and $d=8$ sized Hilbert spaces. We establish and utilize optimization criteria to select qudit computational basis states from the 24 available states in the long-lived, metastable $D_{5/2}$ level. We apply randomized benchmarking sequences for calibration of the operation control parameters. By allowing non-constant pulse lengths, each step of the algorithm is realized with $\mathcal{O}(d)$ pulses, giving us an efficient and high-fidelity implementation. Compared with previous demonstrations of Grover's search algorithm with qubit-based systems, the qudit-based approach provides competitive performance.

\begin{figure}[htp]
    \centering
    \includegraphics[width = 0.42\textwidth]{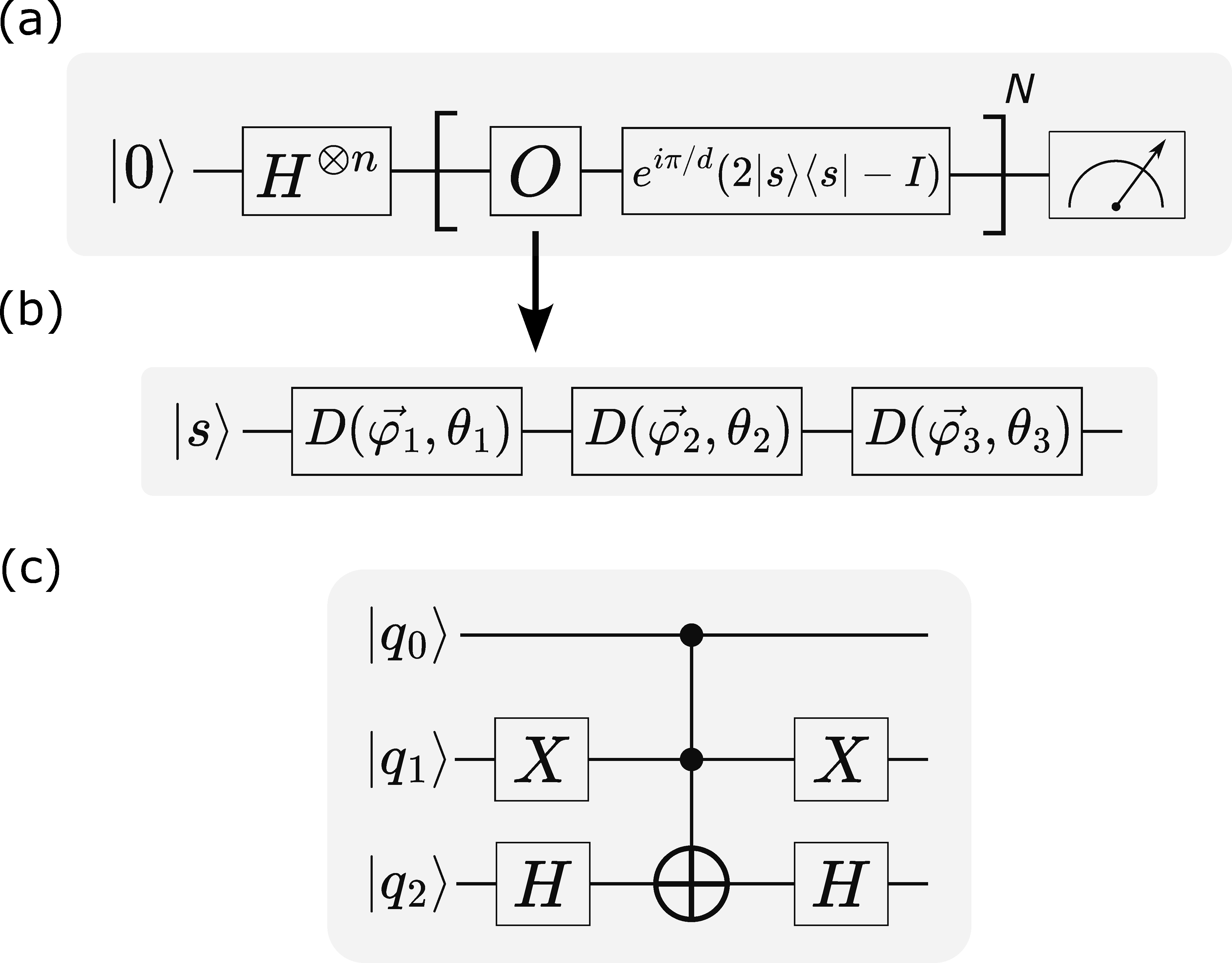}
    \caption{ Grover's algorithm implemented with a single qudit. (a) The qudit is first prepared in the $\ket{0}$ state, followed by a Hardmard transformation that generates the equal superposition state $\ket{s}$. For each iteration, the oracle $O$ is applied to mark the target state, and the reflection operator $2\ket{s}\bra{s}-I$ amplifies the population of the marked state. The phase factor of $e^{i\pi/d}$ is needed if $d$ is even; otherwise, the reflection operator cannot be implemented using SU$(d)$ operations due to its negative determinant. A comparison of two circuits implementing the phase oracle with a $d=8$ qudit (b), which consists of three displacement pulses, and with three qubits (c), which requires a three-qubit Toffoli gate.}
    \label{fig:grover_circuit}
\end{figure}

\begin{figure*}[ht]
    \centering
    \includegraphics[width = 0.95\textwidth]{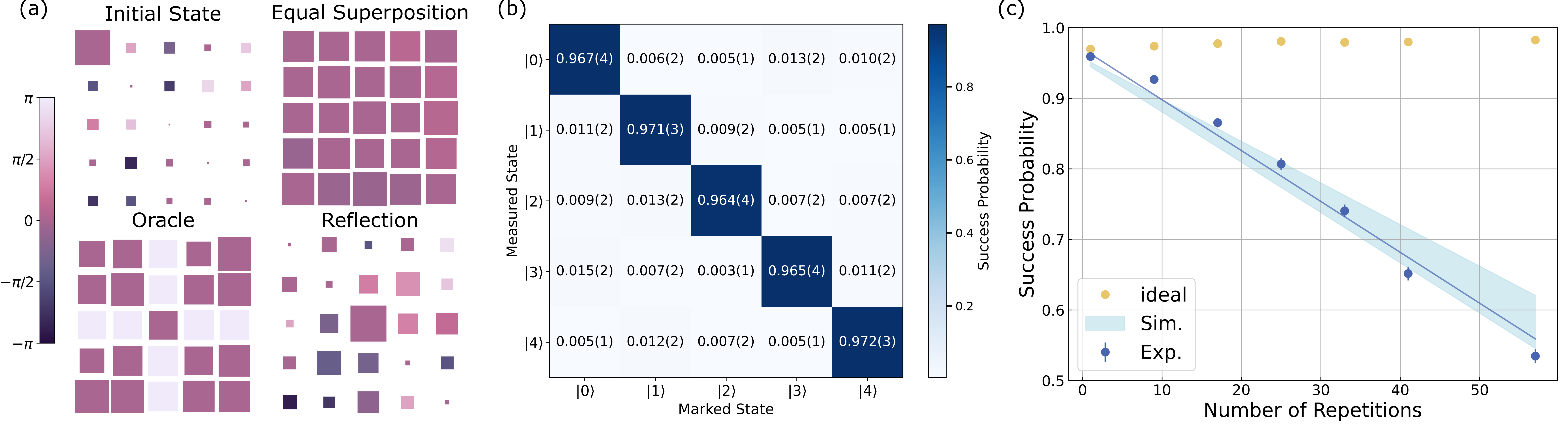}
    \caption{Results for implementation of Grover's algorithm in a $d=5$ qudit. (a) Density matrices reconstructed from state tomography of the qudit after each step of the algorithm. The qudit is initialized in the $\ket{0}$ state, followed by the preparation of the equal superposition state, the oracle operation that marks the $\ket{2}$ state, and the reflection operation that amplifies the amplitude of the marked state. The size (color) of the square blocks represents the magnitude (phase) of the corresponding entry of the density matrix. (b) Result of performing Grover's search, with oracles that mark each qudit state, after one round of oracle-reflection operation. The measurement results are represented by the number within each block. The theoretically achievable fidelity is $\sim$96.7$\%$ and the average fidelity of the experiment is 96.8(3) $\%$, where the error represents one standard deviation. The average SSO is 99.9(1) $\%$ and the highest possible SSO is 100$\%$. (c) The success probability of the algorithm versus the number of repetitions $N$ of the oracle-reflection operations. The yellow and blue markers represent the highest possible fidelity and the experimental result, respectively. The blue solid line represents a linear fit to the data points and the success probability per iteration is determined to be 99.28(2)$\%$. The light shaded-blue region represents the expected fidelity determined from a master-equation-based simulation and the measured coherence time~\cite{supp_material}. } 
    \label{fig:d5_grover}
\end{figure*}

\section{Results}
The experiment is performed with a single \baonethirtyseven\ ion trapped in a surface electrode trap~\cite{Shi2025}. Due to a nuclear spin of $I = \frac{3}{2}$, the metastable \dfh\ state has twenty-four levels available for the construction of qudits. The hyperfine interaction also provides the non-linearity needed for universal control~\cite{champion24, Yu2025}. To obtain states suitable for high-fidelity multi-tone control, we must identify an optimal subset of states coupled to each other through magnetic dipole transitions. These transitions should be relatively strong, with low sensitivity to magnetic field fluctuations, and well separated from any other transition that involves the qudit states or potential leakage paths. Based on these criteria, we numerically vary the strength of the quantization field and select a set of up to eight levels for encoding qudits. We determined a suitable field strength $|\vec{B_{0}}|$ to be $\sim$7.2 Gauss, which is applied through a pair of coils in a near-Helmholtz geometry. The selected states are depicted in Fig.~\ref{fig:levels}(a). A slightly different set of states is chosen for the $d=5$ and the $d=8$ qudits, as we have to relax the constraint somewhat on the separation of the transitions for the latter case. The inter-state transition strength is plotted versus frequency in Fig.\ref{fig:levels}(b) and (c).

\begin{figure}[tp]
    \centering
    \includegraphics[width = 0.5\textwidth]{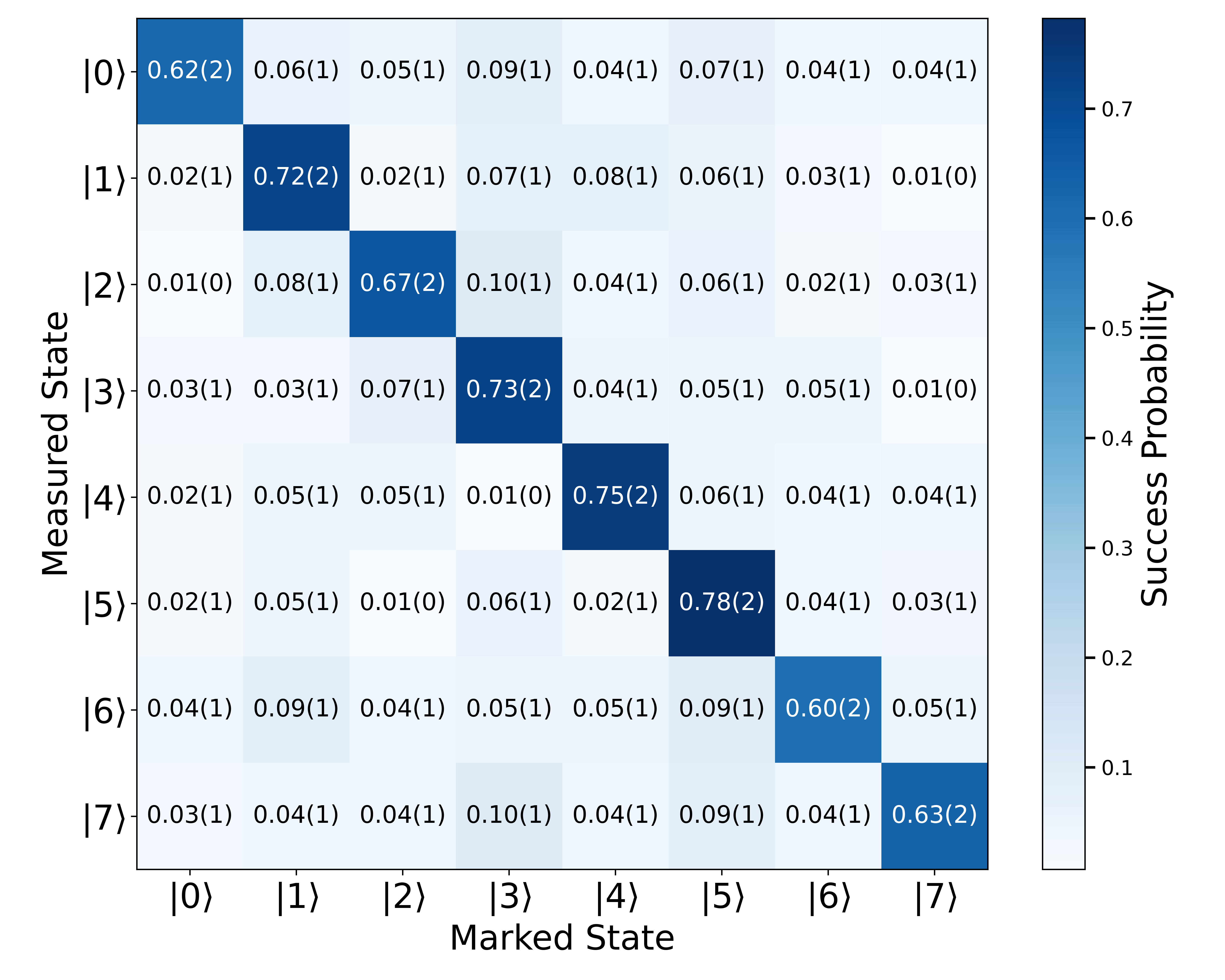}
    \caption{Grover's algorithm implemented on a $d= 8$ qudit. The measured ASP and SSO are 69(6)$\%$ and 97.1(3)$\%$, respectively, while the highest theoretical ASP and SSO are 78$\%$ and 100$\%$, respectively, with one round of oracle-reflection operation. }
    \label{fig:d8_grover}
\end{figure}

A 1762 nm laser beam is applied to drive transitions between the $6S_{1/2}$ and \dfh levels for state preparation and measurement (SPAM)~\cite{An2022}. For qudit states not directly accessible due to selection rules, the population is first transferred to an accessible state via a $\pi$ pulse on an accessible $S$-to-$D$ transition; this is followed by an radio-frequency (RF) $\pi$ pulse that transfers the population to the desired qudit state. A similar procedure is implemented to sequentially transfer the population from the qudit states back to the ground state for measurement via fluorescence. The total time required to completely read out the qudit is ${\sim}8$~ms; this is sufficiently short compared to the lifetime of the \dfh state such that decay during measurement leads to negligible error. We detect leakage out of the qudit state space during operations as follows:  if the ion is never detected to be bright in any of the cyclical qudit-state-transfer measurement cycles, the experimental trial is discarded~\cite{Sotirova2024}.  For the measurement results depicted in Fig.~\ref{fig:d5_grover}(b) and Fig.~\ref{fig:d8_grover}, the average probability of null measurements is $2(1)\%$, limited by the SPAM fidelity. 

To perform multi-tone operations, up to seven RF signals (each generated via direct-digital synthesis [DDS], and all relatively phase-coherent) are combined and directed through two RF electrodes in the trap, as shown in Fig.~\ref{fig:levels}(d). The oscillating magnetic field generated by the RF electrodes has components along both the $x$ and $z$ directions. Therefore, magnetic dipole transitions of any polarization can be driven among the sublevels of the \dfh\ manifold. Concerted, independent control of the multi-tone phases allows for displacement and Selective Number-dependent Arbitrary Phase (SNAP) gates for universal control~\cite{Yu2025}.

For calibration of the multi-tone control, we first perform spectroscopy and Rabi-excitation experiments to locate the resonances of all the relevant transitions and characterize the approximate driving strength for each transition's tone individually. The amplitudes of the DDS channels are adjusted so that the Rabi frequencies approximate those required for a spin-displacement operation (see Methods). However, this coarse calibration is not sufficient to guarantee low gate error, likely because driving all tones simultaneously introduces additional effects, such as AC Zeeman shifts, which are not taken into account at this level of calibration. Hence, coarse calibration is then followed by a finer calibration of the Rabi amplitudes using the Nelder-Mead method to maximize the fidelity of global Clifford gate sequences composed of the native $\pi$-pulses \cite{supp_material}. These sequences are randomized benchmarking sequences and provide an optimization landscape for the multi-tone calibration with a local minimum at the optimal spin-displacement amplitudes as described in the "Methods" section. Averaging many of these sequences creates a global minimum within a reasonable tuning range from the coarse approximation (see \cite{supp_material} for further details).

Grover's search algorithm provides quadratic speed up compared to classical brute-force search when querying an unsorted dataset. The circuit diagram for implementing this algorithm with a single qudit is shown in Fig.~\ref{fig:grover_circuit}. A few straightforward modifications to the textbook Grover's algorithm are needed when implemented with a single qudit, versus $\log_2 d$ qubits.
As in the standard algorithm, we begin by constructing the equal superposition over all states in the database
\begin{align}
    \ket{s} = \frac{1}{\sqrt{d}}\sum_{k=0}^{d-1} \ket{k}.
    \label{equal_sup_eq}
\end{align}
We then implement a phase oracle $O$ which marks the desired state $\ket{m}$ with a phase of $-1$; this is again just like in the qubit case.  On the other hand, in the next step a change needs to be made to the reflection operator
$2\kb{s}{s} - \mathbb{1}$, which is applied to amplify the population in the marked state. In even dimensions of $d>2$, one must scale the reflection operator by a phase factor of $e^{i \pi/d}$ to ensure that the operator belongs to $SU(d)$. The marking and reflection operations are then repeated $N\sim \pi\sqrt{d}/4$ times, after which the algorithm success probability (ASP) is given by ~\cite{Nielsen_Chuang_2010} 
\begin{equation}
    p(N) = \sin^2\left [(2N+1)\sin^{-1} \left(\frac{1}{\sqrt{d}} \right)\right ]. 
    \label{grover_max_fid}
\end{equation}

We implement and characterize the performance of Grover's algorithm as follows. The sequence starts with the preparation of the ion in the $\ket{0}$ state of the qudit, followed by the creation of the equal superposition state as given by Eq.~\ref{equal_sup_eq}, which takes two pulses for $d= 5$. For the oracle operation, two pulses are applied to mark the desired state. Lastly, four pulses are used to implement the reflection operator. (The pulse parameters are summarized in \cite{supp_material}.) We perform state tomography \cite{Hradi1997} to diagnose the density matrix after each stage of the algorithm, and the result is shown in Fig.~\ref{fig:d5_grover}(a). The algorithm is repeated with oracles that mark different qudit states, and this result is shown in Fig.~\ref{fig:d5_grover}(b). An average ASP of 96.8(3)$\%$ is achieved for $N = 1$, while the highest possible ASP is $\sim$96.7$\%$, given by Eq.~\ref{grover_max_fid}.

We further evaluate the success of our experimental implementation using the squared statistical overlap (SSO), a metric that compares the population of each measured state with the theoretical prediction.  The SSO is defined as $\left(\sum_{k=0}^d\sqrt{e_kp_k}\right)^2$, where $e_k$ is the predicted population, and $p_k$ is the measured population, of the $k$-th state. The average SSO for our $d=5$ implementation is 99.9(1)$\%$, while the highest possible SSO is 100$\%$. To more precisely determine the fidelity per round, we perform multiple iterations of the oracle-reflection operation, and the result is shown in Fig.~\ref{fig:d5_grover}(c). The highest possible fidelity is also plotted for reference. A linear fit to the decay of the success probability results in a fidelity of 99.28(2)$\%$ for each repetition. 

To appreciate how well the results scale to larger qudit sizes, we also implement Grover's algorithm with a $d=8$ qudit, and evaluate the required control complexity and the obtained algorithm fidelity. We observe that certain algorithmic operations could be significantly simplified with qudits, as shown in Fig.~\ref{fig:grover_circuit}(b), compared with similarly sized qubit-based systems~\cite{Figgatt2017, AbuGhanem2025, Tanay2020}. The Hardmard transformation, oracle, and reflection operations are implemented with three, two, and eight displacement pulses, respectively. The experimental result with $N=1$ round of oracle-reflection operations is shown in Fig.~\ref{fig:d8_grover}. The average ASP is 69(6)$\%$ and the average SSO is 97.1(3)$\%$, with the highest possible ASP, from Eq.~\ref{grover_max_fid} being 78$\%$, and the highest possible SSO being 100 $\%$.

\section{Discussion}
A major source of error that limits the algorithm's fidelity is decoherence. For the $d= 5$ and $8$ cases studied here, the measured coherence times of the states utilized are 3(1) and 9(1) ms, respectively \cite{supp_material}, while the average pulse lengths for implementing the algorithm are  $33$ and $30$ $\mu s$, respectively. Therefore, we expect decoherence to contribute approximately 0.4$\%$ and 1$\%$ error per pulse, respectively. We also simulate the expected fidelity versus the number of oracle-reflection repetitions for the $d= 5$ case using a master-equation approach, with a dephasing operator that is a diagonal matrix where the matrix elements are the magnetic field sensitivities of each level (assuming the magnetic field fluctuation is small and is the major source of decoherence, which is a good assumption in our system). The result of the simulation is plotted next to the measurements in Fig.~\ref{fig:d5_grover}(c),  showing reasonable agreement and confirming that decoherence is the dominant source of error.  Another source of error that also increases with the qudit dimension is the off-resonant coupling to non-qudit states. The off-resonant drive that couples state $\ket{i}$ and $\ket{k}$ results in a population $\sim\frac{\Omega_{ik}^2}{\delta_{ik}^2}$, where $\Omega_{ik}$ is the coupling strength and $\delta_{ik}$ is the detuning, each between states $i$ and $k$. Summing over all possible couplings when the qudit drives are switched on gives an expected error of $\sim 3\times10^{-5}$ and $\sim 2\times10^{-4}$, respectively.

Grover's search over a data size of eight has also been performed on qubit-based systems, including trapped ions \cite{Figgatt2017} and superconducting circuits \cite{AbuGhanem2025, Tanay2020}. For a single iteration with the phase oracle, the average ASP to identify the marked state was 43.7(2)$\%$, 51(6)$\%$, and 49.2(4)$\%$ for each of these referenced experiments, respectively, approximately 20$\%$ lower than our implementation with a single qudit. We note that error correction is not employed in these implementations and higher fidelity should be possible. As noted earlier, a major challenge for the qubit-based implementation is the large number of two-qubit gates. In our case, no entangling gate is needed to implement Grover's search algorithm for this database size. This reduction of the number of entangling gates could extend to larger database sizes, due to the much more efficient decomposition of qudit Toffoli gates \cite{Amit2022, Nikolaeva2023}. Native qudit entangling gates between trapped ion qudits have already been demonstrated in \cite{Hrmo2023} with a fidelity of 93.7(3) $\%$, limited by dephasing caused by technical noise sources. If qudit entangling gates could be performed with high fidelity and for higher $d$, it might be more favorable to implement the algorithm with qudit-based systems, particularly if qudit-based quantum error correction is also practical and effective.

In conclusion, we have demonstrated efficient, high-fidelity multi-tone control of a single qudit of up to eight states encoded in a single trapped ion. This control scheme allows for the implementation of Grover's search algorithm without entangling gates, and our implementation achieves a fidelity comparable to that obtained in qubit-based systems.

\section{Methods}
\subsubsection{Energy Levels of \ion{Ba}{137} Metastable states}
For atoms with a non-zero nuclear spin I, the Hamiltonian is given by (neglecting higher order contributions)
\begin{align}
    \begin{split}
    H_0 = A\bb{I}\cdot\bb{J} + & B \frac{3(\vec{I}\cdot\vec{J})^2+\frac{3}{2}(\vec{I}\cdot\vec{J}) - I(I+1)(J+1)}{2I(2I-1)J(2J-1)} \\ &+\mu_BB_z(g_j m_J + g_I m_I)
    \end{split}    
\end{align}
where $A$ and $B$ are the magnetic dipole and electric quadrupole hyperfine constants, respectively, $g_J$ and $g_I$ are the Land\'{e} g-factor and the nuclear g-factor. Here, $\vec{I}$ and $\vec{J}$ are the nuclear and total angular momentum operator, and $m_I$,$m_J$ are their projection along the quantization axis. $B_z$ is the strength of the magnetic field along the quantization axis(the z-axis). With a non-zero $B_z$, the eigenstates are linear superpositions of all states with the same $m_F$, which is the component of the total angular momentum $\vec{F} = \vec{I} + \vec{J}$ along the quantization field. We select a good set of states in the \dfh level for the qudit. The states are labeled as $\ket{0}, \ket{1}, ... \ket{d-1}$ and mapped to an effective spin-$\frac{(d-1)}{2}$ with an assigned $J_z$ value for $-\frac{d}{2}+i$ to state $\ket{i}$.

\subsubsection{Multi-tone control}
When the atom interacts with a multi-tone magnetic field, with the $k$-th tone having drive frequency $\omega_{k}$, field strength $B_{k}$, and phase $\varphi_k$, the interaction Hamiltonian $H_{\rm lab}$ in the laboratory frame is given by
\begin{align}
    H_{\rm lab}(t) = \mu_Bg_jJ_x\sum_{k}{B}_k  \cos{(\w_k t + \varphi_k)}
\end{align}
where $J_x$ is the D$_{5/2}$ electronic spin operator. After going into the generalized rotating frame defined by $U = \sum_{k}e^{iE_{k} t/\h} \kb{k}{k}$, where $E_{k}$ is the energy of the $k$th level, and making the rotating-wave approximation, the Hamiltonian becomes time-independent and is given by \cite{leuenberger03}

\begin{equation}
    H_{rot} = 
    \begin{bmatrix}
        0 & \rabi_0e^{i\varphi_0} & 0 & \dots  \\
        \rabi_0e^{-i\varphi_0} & \delta_0 &  \rabi_1e^{i\varphi_1} &\vdots\\
        0 & \rabi_1e^{-i\varphi_1} & \ddots &  \rabi_de^{i\varphi_{d-1}}\\
        \vdots & \dots&  \rabi_de^{-i\varphi_{d-1}}& \sum_{k=0}^{d-2}\delta_k\\
    \end{bmatrix},
\end{equation}

\noindent
where $\rabi_k = \mu_Bg_j{B}_k|\bra{k}\sigma_x^{k}\ket{k+1}|$, $\sigma_x^{k}$ is the Pauli operator that couples state $\ket{k}$ and $\ket{k+1}$, and $\delta_k = (E_{k+1}-E_k) - \omega_k$. When the amplitudes of each tone satisfy
\begin{align}
    \rabi_{k-1} = \Omega \sqrt{k(d-k)}
    \label{rabi_eq}
\end{align}
we have $H_{rot}(\vec{\varphi}=\vec{0}) = \Omega J_x$, where $\Omega$ is the Rabi frequency for the $J_x$ rotation. These multi-tone spin-displacement pulses $D(\vec{\varphi}, \theta) \equiv e^{-i\theta H_{rot}(\vec{\varphi})}$, where $\theta = \Omega t$, offer universal control over the qudit and provide a natural construction for pulse sequences with linear depth scaling in $d$ that approximate arbitrary unitary $U$ in $SU(d)$\cite{champion24}:
\begin{align}
    U \approx \prod_{n=1}^d D(\vec{\varphi}_n, \theta_n)
\end{align}
This scheme is analogous to the SNAP-displacement \cite{champion24} operations where the phases $\vec{\varphi}$ can be understood as the effect of virtual SNAP gates \cite{mckay_efficient_2017} $S(\vec{\varphi})$ interleaved with spin-displacements $D(\theta) = D(\theta, \vec{\varphi}=\vec{0}) $:
\begin{align}
    U = S(\vec{\varphi_0})\prod_{n=1}^d D(\theta_n)S(\vec{\varphi_n})
    \,.
\end{align}

\subsubsection{Algorithm Operation Pulse Sequence}
The core challenge for mapping  Grover's algorithm onto a single qudit is the efficient realization of the $d$-dimensional unitary operators. With a structured universal gate set, an arbitrary unitary could be constructed using $\mathcal{O}(d^2)$ pulses~\cite{Bullock2005}. In contrast, an unstructured gate set (readily available with the multi-tone control) can achieve $\mathcal{O}(d)$ \cite{champion24} scaling, at the cost of straightforward decompositions. To address this challenge, we use a gradient-descent-based numerical optimization method to search for pulse parameters that implement the desired unitary based on a defined ansatz. For example, the marking operations in the phase oracle are implemented using an ansatz consisting of two successive displacement pulses, each defined by the rotation angle and phases of each tone. The optimization method then minimizes a loss function defined by the distance from the unitary generated by the two pulses to the oracle. Short sequences are found in this way for each step of our Grover's algorithm implementation; these sequences are summarized in \cite{supp_material}.

\section*{Acknowledgements}

I.\,L.\,C. acknowledges support by the NSF Center for Ultracold Atoms.  This research was supported by the U.S. Army Research Office through grant W911NF-24-1-0379.  This material is based upon work supported by the Department of Defense under Air Force Contract No. FA8702-15-D-0001. Any opinions, findings, conclusions, or recommendations expressed in this material are those of the author(s) and do not necessarily reflect the views of the Department of Defense.

    

\bibliographystyle{apsrev4-2}
\bibliography{references}

\begin{thebibliography}{48}%
\makeatletter
\providecommand \@ifxundefined [1]{%
 \@ifx{#1\undefined}
}%
\providecommand \@ifnum [1]{%
 \ifnum #1\expandafter \@firstoftwo
 \else \expandafter \@secondoftwo
 \fi
}%
\providecommand \@ifx [1]{%
 \ifx #1\expandafter \@firstoftwo
 \else \expandafter \@secondoftwo
 \fi
}%
\providecommand \natexlab [1]{#1}%
\providecommand \enquote  [1]{``#1''}%
\providecommand \bibnamefont  [1]{#1}%
\providecommand \bibfnamefont [1]{#1}%
\providecommand \citenamefont [1]{#1}%
\providecommand \href@noop [0]{\@secondoftwo}%
\providecommand \href [0]{\begingroup \@sanitize@url \@href}%
\providecommand \@href[1]{\@@startlink{#1}\@@href}%
\providecommand \@@href[1]{\endgroup#1\@@endlink}%
\providecommand \@sanitize@url [0]{\catcode `\\12\catcode `\$12\catcode `\&12\catcode `\#12\catcode `\^12\catcode `\_12\catcode `\%12\relax}%
\providecommand \@@startlink[1]{}%
\providecommand \@@endlink[0]{}%
\providecommand \url  [0]{\begingroup\@sanitize@url \@url }%
\providecommand \@url [1]{\endgroup\@href {#1}{\urlprefix }}%
\providecommand \urlprefix  [0]{URL }%
\providecommand \Eprint [0]{\href }%
\providecommand \doibase [0]{https://doi.org/}%
\providecommand \selectlanguage [0]{\@gobble}%
\providecommand \bibinfo  [0]{\@secondoftwo}%
\providecommand \bibfield  [0]{\@secondoftwo}%
\providecommand \translation [1]{[#1]}%
\providecommand \BibitemOpen [0]{}%
\providecommand \bibitemStop [0]{}%
\providecommand \bibitemNoStop [0]{.\EOS\space}%
\providecommand \EOS [0]{\spacefactor3000\relax}%
\providecommand \BibitemShut  [1]{\csname bibitem#1\endcsname}%
\let\auto@bib@innerbib\@empty
\bibitem [{\citenamefont {Bluvstein}\ \emph {et~al.}(2024)\citenamefont {Bluvstein}, \citenamefont {Evered}, \citenamefont {Geim}, \citenamefont {Li}, \citenamefont {Zhou}, \citenamefont {Manovitz}, \citenamefont {Ebadi}, \citenamefont {Cain}, \citenamefont {Kalinowski}, \citenamefont {Hangleiter}, \citenamefont {Bonilla~Ataides}, \citenamefont {Maskara}, \citenamefont {Cong}, \citenamefont {Gao}, \citenamefont {Sales~Rodriguez}, \citenamefont {Karolyshyn}, \citenamefont {Semeghini}, \citenamefont {Gullans}, \citenamefont {Greiner}, \citenamefont {Vuleti{\'{c}}},\ and\ \citenamefont {Lukin}}]{Bluvstein2024}%
  \BibitemOpen
  \bibfield  {author} {\bibinfo {author} {\bibfnamefont {D.}~\bibnamefont {Bluvstein}}, \bibinfo {author} {\bibfnamefont {S.~J.}\ \bibnamefont {Evered}}, \bibinfo {author} {\bibfnamefont {A.~A.}\ \bibnamefont {Geim}}, \bibinfo {author} {\bibfnamefont {S.~H.}\ \bibnamefont {Li}}, \bibinfo {author} {\bibfnamefont {H.}~\bibnamefont {Zhou}}, \bibinfo {author} {\bibfnamefont {T.}~\bibnamefont {Manovitz}}, \bibinfo {author} {\bibfnamefont {S.}~\bibnamefont {Ebadi}}, \bibinfo {author} {\bibfnamefont {M.}~\bibnamefont {Cain}}, \bibinfo {author} {\bibfnamefont {M.}~\bibnamefont {Kalinowski}}, \bibinfo {author} {\bibfnamefont {D.}~\bibnamefont {Hangleiter}}, \bibinfo {author} {\bibfnamefont {J.~P.}\ \bibnamefont {Bonilla~Ataides}}, \bibinfo {author} {\bibfnamefont {N.}~\bibnamefont {Maskara}}, \bibinfo {author} {\bibfnamefont {I.}~\bibnamefont {Cong}}, \bibinfo {author} {\bibfnamefont {X.}~\bibnamefont {Gao}}, \bibinfo {author} {\bibfnamefont {P.}~\bibnamefont {Sales~Rodriguez}}, \bibinfo {author} {\bibfnamefont
  {T.}~\bibnamefont {Karolyshyn}}, \bibinfo {author} {\bibfnamefont {G.}~\bibnamefont {Semeghini}}, \bibinfo {author} {\bibfnamefont {M.~J.}\ \bibnamefont {Gullans}}, \bibinfo {author} {\bibfnamefont {M.}~\bibnamefont {Greiner}}, \bibinfo {author} {\bibfnamefont {V.}~\bibnamefont {Vuleti{\'{c}}}},\ and\ \bibinfo {author} {\bibfnamefont {M.~D.}\ \bibnamefont {Lukin}},\ }\href {https://doi.org/10.1038/s41586-023-06927-3} {\bibfield  {journal} {\bibinfo  {journal} {Nature}\ }\textbf {\bibinfo {volume} {626}},\ \bibinfo {pages} {58} (\bibinfo {year} {2024})}\BibitemShut {NoStop}%
\bibitem [{\citenamefont {Manetsch}\ \emph {et~al.}(2024)\citenamefont {Manetsch}, \citenamefont {Nomura}, \citenamefont {Bataille}, \citenamefont {Leung}, \citenamefont {Lv},\ and\ \citenamefont {Endres}}]{Manetsch2024}%
  \BibitemOpen
  \bibfield  {author} {\bibinfo {author} {\bibfnamefont {H.~J.}\ \bibnamefont {Manetsch}}, \bibinfo {author} {\bibfnamefont {G.}~\bibnamefont {Nomura}}, \bibinfo {author} {\bibfnamefont {E.}~\bibnamefont {Bataille}}, \bibinfo {author} {\bibfnamefont {K.~H.}\ \bibnamefont {Leung}}, \bibinfo {author} {\bibfnamefont {X.}~\bibnamefont {Lv}},\ and\ \bibinfo {author} {\bibfnamefont {M.}~\bibnamefont {Endres}},\ }\href@noop {} {\bibfield  {journal} {\bibinfo  {journal} {arXiv preprint arXiv:2403.12021}\ } (\bibinfo {year} {2024})}\BibitemShut {NoStop}%
\bibitem [{\citenamefont {Bruzewicz}\ \emph {et~al.}(2019)\citenamefont {Bruzewicz}, \citenamefont {Chiaverini}, \citenamefont {McConnell},\ and\ \citenamefont {Sage}}]{Bruzewicz2019}%
  \BibitemOpen
  \bibfield  {author} {\bibinfo {author} {\bibfnamefont {C.~D.}\ \bibnamefont {Bruzewicz}}, \bibinfo {author} {\bibfnamefont {J.}~\bibnamefont {Chiaverini}}, \bibinfo {author} {\bibfnamefont {R.}~\bibnamefont {McConnell}},\ and\ \bibinfo {author} {\bibfnamefont {J.~M.}\ \bibnamefont {Sage}},\ }\href@noop {} {\bibfield  {journal} {\bibinfo  {journal} {Applied physics reviews}\ }\textbf {\bibinfo {volume} {6}} (\bibinfo {year} {2019})}\BibitemShut {NoStop}%
\bibitem [{\citenamefont {Henriet}\ \emph {et~al.}(2020)\citenamefont {Henriet}, \citenamefont {Beguin}, \citenamefont {Signoles}, \citenamefont {Lahaye}, \citenamefont {Browaeys}, \citenamefont {Reymond},\ and\ \citenamefont {Jurczak}}]{Henriet2020}%
  \BibitemOpen
  \bibfield  {author} {\bibinfo {author} {\bibfnamefont {L.}~\bibnamefont {Henriet}}, \bibinfo {author} {\bibfnamefont {L.}~\bibnamefont {Beguin}}, \bibinfo {author} {\bibfnamefont {A.}~\bibnamefont {Signoles}}, \bibinfo {author} {\bibfnamefont {T.}~\bibnamefont {Lahaye}}, \bibinfo {author} {\bibfnamefont {A.}~\bibnamefont {Browaeys}}, \bibinfo {author} {\bibfnamefont {G.-O.}\ \bibnamefont {Reymond}},\ and\ \bibinfo {author} {\bibfnamefont {C.}~\bibnamefont {Jurczak}},\ }\href@noop {} {\bibfield  {journal} {\bibinfo  {journal} {Quantum}\ }\textbf {\bibinfo {volume} {4}},\ \bibinfo {pages} {327} (\bibinfo {year} {2020})}\BibitemShut {NoStop}%
\bibitem [{\citenamefont {Krantz}\ \emph {et~al.}(2019)\citenamefont {Krantz}, \citenamefont {Kjaergaard}, \citenamefont {Yan}, \citenamefont {Orlando}, \citenamefont {Gustavsson},\ and\ \citenamefont {Oliver}}]{Krantz2019}%
  \BibitemOpen
  \bibfield  {author} {\bibinfo {author} {\bibfnamefont {P.}~\bibnamefont {Krantz}}, \bibinfo {author} {\bibfnamefont {M.}~\bibnamefont {Kjaergaard}}, \bibinfo {author} {\bibfnamefont {F.}~\bibnamefont {Yan}}, \bibinfo {author} {\bibfnamefont {T.~P.}\ \bibnamefont {Orlando}}, \bibinfo {author} {\bibfnamefont {S.}~\bibnamefont {Gustavsson}},\ and\ \bibinfo {author} {\bibfnamefont {W.~D.}\ \bibnamefont {Oliver}},\ }\href@noop {} {\bibfield  {journal} {\bibinfo  {journal} {Applied physics reviews}\ }\textbf {\bibinfo {volume} {6}} (\bibinfo {year} {2019})}\BibitemShut {NoStop}%
\bibitem [{\citenamefont {Monz}\ \emph {et~al.}(2016)\citenamefont {Monz}, \citenamefont {Nigg}, \citenamefont {Martinez}, \citenamefont {Brandl}, \citenamefont {Schindler}, \citenamefont {Rines}, \citenamefont {Wang}, \citenamefont {Chuang},\ and\ \citenamefont {Blatt}}]{Monz2016}%
  \BibitemOpen
  \bibfield  {author} {\bibinfo {author} {\bibfnamefont {T.}~\bibnamefont {Monz}}, \bibinfo {author} {\bibfnamefont {D.}~\bibnamefont {Nigg}}, \bibinfo {author} {\bibfnamefont {E.~A.}\ \bibnamefont {Martinez}}, \bibinfo {author} {\bibfnamefont {M.~F.}\ \bibnamefont {Brandl}}, \bibinfo {author} {\bibfnamefont {P.}~\bibnamefont {Schindler}}, \bibinfo {author} {\bibfnamefont {R.}~\bibnamefont {Rines}}, \bibinfo {author} {\bibfnamefont {S.~X.}\ \bibnamefont {Wang}}, \bibinfo {author} {\bibfnamefont {I.~L.}\ \bibnamefont {Chuang}},\ and\ \bibinfo {author} {\bibfnamefont {R.}~\bibnamefont {Blatt}},\ }\href@noop {} {\bibfield  {journal} {\bibinfo  {journal} {Science}\ }\textbf {\bibinfo {volume} {351}},\ \bibinfo {pages} {1068} (\bibinfo {year} {2016})}\BibitemShut {NoStop}%
\bibitem [{\citenamefont {Figgatt}\ \emph {et~al.}(2017)\citenamefont {Figgatt}, \citenamefont {Maslov}, \citenamefont {Landsman}, \citenamefont {Linke}, \citenamefont {Debnath},\ and\ \citenamefont {Monroe}}]{Figgatt2017}%
  \BibitemOpen
  \bibfield  {author} {\bibinfo {author} {\bibfnamefont {C.}~\bibnamefont {Figgatt}}, \bibinfo {author} {\bibfnamefont {D.}~\bibnamefont {Maslov}}, \bibinfo {author} {\bibfnamefont {K.~A.}\ \bibnamefont {Landsman}}, \bibinfo {author} {\bibfnamefont {N.~M.}\ \bibnamefont {Linke}}, \bibinfo {author} {\bibfnamefont {S.}~\bibnamefont {Debnath}},\ and\ \bibinfo {author} {\bibfnamefont {C.}~\bibnamefont {Monroe}},\ }\href {https://doi.org/10.1038/s41467-017-01904-7} {\bibfield  {journal} {\bibinfo  {journal} {Nature Communications}\ }\textbf {\bibinfo {volume} {8}},\ \bibinfo {pages} {1918} (\bibinfo {year} {2017})}\BibitemShut {NoStop}%
\bibitem [{\citenamefont {Barenco}\ \emph {et~al.}(1995)\citenamefont {Barenco}, \citenamefont {Bennett}, \citenamefont {Cleve}, \citenamefont {DiVincenzo}, \citenamefont {Margolus}, \citenamefont {Shor}, \citenamefont {Sleator}, \citenamefont {Smolin},\ and\ \citenamefont {Weinfurter}}]{Adriano1995}%
  \BibitemOpen
  \bibfield  {author} {\bibinfo {author} {\bibfnamefont {A.}~\bibnamefont {Barenco}}, \bibinfo {author} {\bibfnamefont {C.~H.}\ \bibnamefont {Bennett}}, \bibinfo {author} {\bibfnamefont {R.}~\bibnamefont {Cleve}}, \bibinfo {author} {\bibfnamefont {D.~P.}\ \bibnamefont {DiVincenzo}}, \bibinfo {author} {\bibfnamefont {N.}~\bibnamefont {Margolus}}, \bibinfo {author} {\bibfnamefont {P.}~\bibnamefont {Shor}}, \bibinfo {author} {\bibfnamefont {T.}~\bibnamefont {Sleator}}, \bibinfo {author} {\bibfnamefont {J.~A.}\ \bibnamefont {Smolin}},\ and\ \bibinfo {author} {\bibfnamefont {H.}~\bibnamefont {Weinfurter}},\ }\href {https://doi.org/10.1103/PhysRevA.52.3457} {\bibfield  {journal} {\bibinfo  {journal} {Phys. Rev. A}\ }\textbf {\bibinfo {volume} {52}},\ \bibinfo {pages} {3457} (\bibinfo {year} {1995})}\BibitemShut {NoStop}%
\bibitem [{\citenamefont {Nielsen}\ and\ \citenamefont {Chuang}(2010)}]{Nielsen_Chuang_2010}%
  \BibitemOpen
  \bibfield  {author} {\bibinfo {author} {\bibfnamefont {M.~A.}\ \bibnamefont {Nielsen}}\ and\ \bibinfo {author} {\bibfnamefont {I.~L.}\ \bibnamefont {Chuang}},\ }\href@noop {} {\emph {\bibinfo {title} {Quantum Computation and Quantum information}}}\ (\bibinfo  {publisher} {Cambridge Univ. Press},\ \bibinfo {year} {2010})\BibitemShut {NoStop}%
\bibitem [{\citenamefont {AbuGhanem}(2025)}]{AbuGhanem2025}%
  \BibitemOpen
  \bibfield  {author} {\bibinfo {author} {\bibfnamefont {M.}~\bibnamefont {AbuGhanem}},\ }\href {https://doi.org/10.1038/s41598-024-80188-6} {\bibfield  {journal} {\bibinfo  {journal} {Scientific Reports}\ }\textbf {\bibinfo {volume} {15}},\ \bibinfo {pages} {1281} (\bibinfo {year} {2025})}\BibitemShut {NoStop}%
\bibitem [{\citenamefont {Nikolaeva}\ \emph {et~al.}(2023)\citenamefont {Nikolaeva}, \citenamefont {Kiktenko},\ and\ \citenamefont {Fedorov}}]{Nikolaeva2023}%
  \BibitemOpen
  \bibfield  {author} {\bibinfo {author} {\bibfnamefont {A.~S.}\ \bibnamefont {Nikolaeva}}, \bibinfo {author} {\bibfnamefont {E.~O.}\ \bibnamefont {Kiktenko}},\ and\ \bibinfo {author} {\bibfnamefont {A.~K.}\ \bibnamefont {Fedorov}},\ }\href@noop {} {\bibfield  {journal} {\bibinfo  {journal} {Entropy}\ }\textbf {\bibinfo {volume} {25}},\ \bibinfo {pages} {387} (\bibinfo {year} {2023})}\BibitemShut {NoStop}%
\bibitem [{\citenamefont {Saeedi}\ and\ \citenamefont {Pedram}(2013)}]{Saeedi2013}%
  \BibitemOpen
  \bibfield  {author} {\bibinfo {author} {\bibfnamefont {M.}~\bibnamefont {Saeedi}}\ and\ \bibinfo {author} {\bibfnamefont {M.}~\bibnamefont {Pedram}},\ }\href@noop {} {\bibfield  {journal} {\bibinfo  {journal} {Physical Review A—Atomic, Molecular, and Optical Physics}\ }\textbf {\bibinfo {volume} {87}},\ \bibinfo {pages} {062318} (\bibinfo {year} {2013})}\BibitemShut {NoStop}%
\bibitem [{\citenamefont {Lubinski}\ \emph {et~al.}(2023)\citenamefont {Lubinski}, \citenamefont {Johri}, \citenamefont {Varosy}, \citenamefont {Coleman}, \citenamefont {Zhao}, \citenamefont {Necaise}, \citenamefont {Baldwin}, \citenamefont {Mayer},\ and\ \citenamefont {Proctor}}]{Lubinski2023}%
  \BibitemOpen
  \bibfield  {author} {\bibinfo {author} {\bibfnamefont {T.}~\bibnamefont {Lubinski}}, \bibinfo {author} {\bibfnamefont {S.}~\bibnamefont {Johri}}, \bibinfo {author} {\bibfnamefont {P.}~\bibnamefont {Varosy}}, \bibinfo {author} {\bibfnamefont {J.}~\bibnamefont {Coleman}}, \bibinfo {author} {\bibfnamefont {L.}~\bibnamefont {Zhao}}, \bibinfo {author} {\bibfnamefont {J.}~\bibnamefont {Necaise}}, \bibinfo {author} {\bibfnamefont {C.~H.}\ \bibnamefont {Baldwin}}, \bibinfo {author} {\bibfnamefont {K.}~\bibnamefont {Mayer}},\ and\ \bibinfo {author} {\bibfnamefont {T.}~\bibnamefont {Proctor}},\ }\href@noop {} {\bibfield  {journal} {\bibinfo  {journal} {IEEE Transactions on Quantum Engineering}\ }\textbf {\bibinfo {volume} {4}},\ \bibinfo {pages} {1} (\bibinfo {year} {2023})}\BibitemShut {NoStop}%
\bibitem [{\citenamefont {Muthukrishnan}\ and\ \citenamefont {Stroud}(2000)}]{Ashok2000}%
  \BibitemOpen
  \bibfield  {author} {\bibinfo {author} {\bibfnamefont {A.}~\bibnamefont {Muthukrishnan}}\ and\ \bibinfo {author} {\bibfnamefont {C.~R.}\ \bibnamefont {Stroud}},\ }\href {https://doi.org/10.1103/PhysRevA.62.052309} {\bibfield  {journal} {\bibinfo  {journal} {Phys. Rev. A}\ }\textbf {\bibinfo {volume} {62}},\ \bibinfo {pages} {052309} (\bibinfo {year} {2000})}\BibitemShut {NoStop}%
\bibitem [{\citenamefont {Ivanov}\ \emph {et~al.}(2012)\citenamefont {Ivanov}, \citenamefont {Tonchev},\ and\ \citenamefont {Vitanov}}]{Ivanov2012}%
  \BibitemOpen
  \bibfield  {author} {\bibinfo {author} {\bibfnamefont {S.~S.}\ \bibnamefont {Ivanov}}, \bibinfo {author} {\bibfnamefont {H.~S.}\ \bibnamefont {Tonchev}},\ and\ \bibinfo {author} {\bibfnamefont {N.~V.}\ \bibnamefont {Vitanov}},\ }\href {https://doi.org/10.1103/PhysRevA.85.062321} {\bibfield  {journal} {\bibinfo  {journal} {Phys. Rev. A}\ }\textbf {\bibinfo {volume} {85}},\ \bibinfo {pages} {062321} (\bibinfo {year} {2012})}\BibitemShut {NoStop}%
\bibitem [{\citenamefont {Saha}\ \emph {et~al.}(2022)\citenamefont {Saha}, \citenamefont {Majumdar}, \citenamefont {Saha}, \citenamefont {Chakrabarti},\ and\ \citenamefont {Sur-Kolay}}]{Amit2022}%
  \BibitemOpen
  \bibfield  {author} {\bibinfo {author} {\bibfnamefont {A.}~\bibnamefont {Saha}}, \bibinfo {author} {\bibfnamefont {R.}~\bibnamefont {Majumdar}}, \bibinfo {author} {\bibfnamefont {D.}~\bibnamefont {Saha}}, \bibinfo {author} {\bibfnamefont {A.}~\bibnamefont {Chakrabarti}},\ and\ \bibinfo {author} {\bibfnamefont {S.}~\bibnamefont {Sur-Kolay}},\ }\href {https://doi.org/10.1103/PhysRevA.105.062453} {\bibfield  {journal} {\bibinfo  {journal} {Phys. Rev. A}\ }\textbf {\bibinfo {volume} {105}},\ \bibinfo {pages} {062453} (\bibinfo {year} {2022})}\BibitemShut {NoStop}%
\bibitem [{\citenamefont {Kiktenko}\ \emph {et~al.}(2020)\citenamefont {Kiktenko}, \citenamefont {Nikolaeva}, \citenamefont {Xu}, \citenamefont {Shlyapnikov},\ and\ \citenamefont {Fedorov}}]{Kiktenko2020}%
  \BibitemOpen
  \bibfield  {author} {\bibinfo {author} {\bibfnamefont {E.~O.}\ \bibnamefont {Kiktenko}}, \bibinfo {author} {\bibfnamefont {A.~S.}\ \bibnamefont {Nikolaeva}}, \bibinfo {author} {\bibfnamefont {P.}~\bibnamefont {Xu}}, \bibinfo {author} {\bibfnamefont {G.~V.}\ \bibnamefont {Shlyapnikov}},\ and\ \bibinfo {author} {\bibfnamefont {A.~K.}\ \bibnamefont {Fedorov}},\ }\href {https://doi.org/10.1103/PhysRevA.101.022304} {\bibfield  {journal} {\bibinfo  {journal} {Phys. Rev. A}\ }\textbf {\bibinfo {volume} {101}},\ \bibinfo {pages} {022304} (\bibinfo {year} {2020})}\BibitemShut {NoStop}%
\bibitem [{\citenamefont {DeBry}\ \emph {et~al.}(2025)\citenamefont {DeBry}, \citenamefont {Meister}, \citenamefont {Martinez}, \citenamefont {Bruzewicz}, \citenamefont {Shi}, \citenamefont {Reens}, \citenamefont {McConnell}, \citenamefont {Chuang},\ and\ \citenamefont {Chiaverini}}]{Debry2025}%
  \BibitemOpen
  \bibfield  {author} {\bibinfo {author} {\bibfnamefont {K.}~\bibnamefont {DeBry}}, \bibinfo {author} {\bibfnamefont {N.}~\bibnamefont {Meister}}, \bibinfo {author} {\bibfnamefont {A.~V.}\ \bibnamefont {Martinez}}, \bibinfo {author} {\bibfnamefont {C.~D.}\ \bibnamefont {Bruzewicz}}, \bibinfo {author} {\bibfnamefont {X.}~\bibnamefont {Shi}}, \bibinfo {author} {\bibfnamefont {D.}~\bibnamefont {Reens}}, \bibinfo {author} {\bibfnamefont {R.}~\bibnamefont {McConnell}}, \bibinfo {author} {\bibfnamefont {I.~L.}\ \bibnamefont {Chuang}},\ and\ \bibinfo {author} {\bibfnamefont {J.}~\bibnamefont {Chiaverini}},\ }\href@noop {} {\bibfield  {journal} {\bibinfo  {journal} {arXiv preprint arXiv:2503.13908}\ } (\bibinfo {year} {2025})}\BibitemShut {NoStop}%
\bibitem [{\citenamefont {Li}\ \emph {et~al.}(2025)\citenamefont {Li}, \citenamefont {Mei}, \citenamefont {Jie}, \citenamefont {Cai}, \citenamefont {Li}, \citenamefont {Liu}, \citenamefont {Chen}, \citenamefont {Xie}, \citenamefont {Cheng}, \citenamefont {Zhao} \emph {et~al.}}]{Li2025}%
  \BibitemOpen
  \bibfield  {author} {\bibinfo {author} {\bibfnamefont {Y.}~\bibnamefont {Li}}, \bibinfo {author} {\bibfnamefont {Q.}~\bibnamefont {Mei}}, \bibinfo {author} {\bibfnamefont {Q.-X.}\ \bibnamefont {Jie}}, \bibinfo {author} {\bibfnamefont {W.}~\bibnamefont {Cai}}, \bibinfo {author} {\bibfnamefont {Y.}~\bibnamefont {Li}}, \bibinfo {author} {\bibfnamefont {Z.}~\bibnamefont {Liu}}, \bibinfo {author} {\bibfnamefont {Z.-J.}\ \bibnamefont {Chen}}, \bibinfo {author} {\bibfnamefont {Z.}~\bibnamefont {Xie}}, \bibinfo {author} {\bibfnamefont {X.}~\bibnamefont {Cheng}}, \bibinfo {author} {\bibfnamefont {X.}~\bibnamefont {Zhao}}, \emph {et~al.},\ }\href@noop {} {\bibfield  {journal} {\bibinfo  {journal} {arXiv preprint arXiv:2504.16746}\ } (\bibinfo {year} {2025})}\BibitemShut {NoStop}%
\bibitem [{\citenamefont {Chiesa}\ \emph {et~al.}(2020)\citenamefont {Chiesa}, \citenamefont {Macaluso}, \citenamefont {Petiziol}, \citenamefont {Wimberger}, \citenamefont {Santini},\ and\ \citenamefont {Carretta}}]{Chiesa2020}%
  \BibitemOpen
  \bibfield  {author} {\bibinfo {author} {\bibfnamefont {A.}~\bibnamefont {Chiesa}}, \bibinfo {author} {\bibfnamefont {E.}~\bibnamefont {Macaluso}}, \bibinfo {author} {\bibfnamefont {F.}~\bibnamefont {Petiziol}}, \bibinfo {author} {\bibfnamefont {S.}~\bibnamefont {Wimberger}}, \bibinfo {author} {\bibfnamefont {P.}~\bibnamefont {Santini}},\ and\ \bibinfo {author} {\bibfnamefont {S.}~\bibnamefont {Carretta}},\ }\href@noop {} {\bibfield  {journal} {\bibinfo  {journal} {The journal of physical chemistry letters}\ }\textbf {\bibinfo {volume} {11}},\ \bibinfo {pages} {8610} (\bibinfo {year} {2020})}\BibitemShut {NoStop}%
\bibitem [{\citenamefont {Ma}\ \emph {et~al.}(2023)\citenamefont {Ma}, \citenamefont {Liu}, \citenamefont {Peng}, \citenamefont {Zhang}, \citenamefont {Jandura}, \citenamefont {Claes}, \citenamefont {Burgers}, \citenamefont {Pupillo}, \citenamefont {Puri},\ and\ \citenamefont {Thompson}}]{Ma2023}%
  \BibitemOpen
  \bibfield  {author} {\bibinfo {author} {\bibfnamefont {S.}~\bibnamefont {Ma}}, \bibinfo {author} {\bibfnamefont {G.}~\bibnamefont {Liu}}, \bibinfo {author} {\bibfnamefont {P.}~\bibnamefont {Peng}}, \bibinfo {author} {\bibfnamefont {B.}~\bibnamefont {Zhang}}, \bibinfo {author} {\bibfnamefont {S.}~\bibnamefont {Jandura}}, \bibinfo {author} {\bibfnamefont {J.}~\bibnamefont {Claes}}, \bibinfo {author} {\bibfnamefont {A.~P.}\ \bibnamefont {Burgers}}, \bibinfo {author} {\bibfnamefont {G.}~\bibnamefont {Pupillo}}, \bibinfo {author} {\bibfnamefont {S.}~\bibnamefont {Puri}},\ and\ \bibinfo {author} {\bibfnamefont {J.~D.}\ \bibnamefont {Thompson}},\ }\href@noop {} {\bibfield  {journal} {\bibinfo  {journal} {Nature}\ }\textbf {\bibinfo {volume} {622}},\ \bibinfo {pages} {279} (\bibinfo {year} {2023})}\BibitemShut {NoStop}%
\bibitem [{\citenamefont {Omanakuttan}\ \emph {et~al.}(2023)\citenamefont {Omanakuttan}, \citenamefont {Mitra}, \citenamefont {Meier}, \citenamefont {Martin},\ and\ \citenamefont {Deutsch}}]{Omanakuttan2023}%
  \BibitemOpen
  \bibfield  {author} {\bibinfo {author} {\bibfnamefont {S.}~\bibnamefont {Omanakuttan}}, \bibinfo {author} {\bibfnamefont {A.}~\bibnamefont {Mitra}}, \bibinfo {author} {\bibfnamefont {E.~J.}\ \bibnamefont {Meier}}, \bibinfo {author} {\bibfnamefont {M.~J.}\ \bibnamefont {Martin}},\ and\ \bibinfo {author} {\bibfnamefont {I.~H.}\ \bibnamefont {Deutsch}},\ }\href {https://doi.org/10.1103/PRXQuantum.4.040333} {\bibfield  {journal} {\bibinfo  {journal} {PRX Quantum}\ }\textbf {\bibinfo {volume} {4}},\ \bibinfo {pages} {040333} (\bibinfo {year} {2023})}\BibitemShut {NoStop}%
\bibitem [{\citenamefont {Omanakuttan}\ \emph {et~al.}(2024)\citenamefont {Omanakuttan}, \citenamefont {Buchemmavari}, \citenamefont {Gross}, \citenamefont {Deutsch},\ and\ \citenamefont {Marvian}}]{Omanakuttan2024}%
  \BibitemOpen
  \bibfield  {author} {\bibinfo {author} {\bibfnamefont {S.}~\bibnamefont {Omanakuttan}}, \bibinfo {author} {\bibfnamefont {V.}~\bibnamefont {Buchemmavari}}, \bibinfo {author} {\bibfnamefont {J.~A.}\ \bibnamefont {Gross}}, \bibinfo {author} {\bibfnamefont {I.~H.}\ \bibnamefont {Deutsch}},\ and\ \bibinfo {author} {\bibfnamefont {M.}~\bibnamefont {Marvian}},\ }\href {https://doi.org/10.1103/PRXQuantum.5.020355} {\bibfield  {journal} {\bibinfo  {journal} {PRX Quantum}\ }\textbf {\bibinfo {volume} {5}},\ \bibinfo {pages} {020355} (\bibinfo {year} {2024})}\BibitemShut {NoStop}%
\bibitem [{\citenamefont {Yu}\ \emph {et~al.}(2025)\citenamefont {Yu}, \citenamefont {Wilhelm}, \citenamefont {Holmes}, \citenamefont {Vaartjes}, \citenamefont {Schwienbacher}, \citenamefont {Nurizzo}, \citenamefont {Kringh{\o}j}, \citenamefont {Blankenstein}, \citenamefont {Jakob}, \citenamefont {Gupta}, \citenamefont {Hudson}, \citenamefont {Itoh}, \citenamefont {Murray}, \citenamefont {Blume-Kohout}, \citenamefont {Ladd}, \citenamefont {Anand}, \citenamefont {Dzurak}, \citenamefont {Sanders}, \citenamefont {Jamieson},\ and\ \citenamefont {Morello}}]{Yu2025}%
  \BibitemOpen
  \bibfield  {author} {\bibinfo {author} {\bibfnamefont {X.}~\bibnamefont {Yu}}, \bibinfo {author} {\bibfnamefont {B.}~\bibnamefont {Wilhelm}}, \bibinfo {author} {\bibfnamefont {D.}~\bibnamefont {Holmes}}, \bibinfo {author} {\bibfnamefont {A.}~\bibnamefont {Vaartjes}}, \bibinfo {author} {\bibfnamefont {D.}~\bibnamefont {Schwienbacher}}, \bibinfo {author} {\bibfnamefont {M.}~\bibnamefont {Nurizzo}}, \bibinfo {author} {\bibfnamefont {A.}~\bibnamefont {Kringh{\o}j}}, \bibinfo {author} {\bibfnamefont {M.~R.~v.}\ \bibnamefont {Blankenstein}}, \bibinfo {author} {\bibfnamefont {A.~M.}\ \bibnamefont {Jakob}}, \bibinfo {author} {\bibfnamefont {P.}~\bibnamefont {Gupta}}, \bibinfo {author} {\bibfnamefont {F.~E.}\ \bibnamefont {Hudson}}, \bibinfo {author} {\bibfnamefont {K.~M.}\ \bibnamefont {Itoh}}, \bibinfo {author} {\bibfnamefont {R.~J.}\ \bibnamefont {Murray}}, \bibinfo {author} {\bibfnamefont {R.}~\bibnamefont {Blume-Kohout}}, \bibinfo {author} {\bibfnamefont {T.~D.}\ \bibnamefont {Ladd}}, \bibinfo {author}
  {\bibfnamefont {N.}~\bibnamefont {Anand}}, \bibinfo {author} {\bibfnamefont {A.~S.}\ \bibnamefont {Dzurak}}, \bibinfo {author} {\bibfnamefont {B.~C.}\ \bibnamefont {Sanders}}, \bibinfo {author} {\bibfnamefont {D.~N.}\ \bibnamefont {Jamieson}},\ and\ \bibinfo {author} {\bibfnamefont {A.}~\bibnamefont {Morello}},\ }\href {https://doi.org/10.1038/s41567-024-02745-0} {\bibfield  {journal} {\bibinfo  {journal} {Nature Physics}\ }\textbf {\bibinfo {volume} {21}},\ \bibinfo {pages} {362} (\bibinfo {year} {2025})}\BibitemShut {NoStop}%
\bibitem [{\citenamefont {Fernández~de Fuentes}\ \emph {et~al.}(2024)\citenamefont {Fernández~de Fuentes}, \citenamefont {Botzem}, \citenamefont {Johnson}, \citenamefont {Vaartjes}, \citenamefont {Asaad}, \citenamefont {Mourik}, \citenamefont {Hudson}, \citenamefont {Itoh}, \citenamefont {Johnson}, \citenamefont {Jakob}, \citenamefont {McCallum}, \citenamefont {Jamieson}, \citenamefont {Dzurak},\ and\ \citenamefont {Morello}}]{morello24}%
  \BibitemOpen
  \bibfield  {author} {\bibinfo {author} {\bibfnamefont {I.}~\bibnamefont {Fernández~de Fuentes}}, \bibinfo {author} {\bibfnamefont {T.}~\bibnamefont {Botzem}}, \bibinfo {author} {\bibfnamefont {M.~A.~I.}\ \bibnamefont {Johnson}}, \bibinfo {author} {\bibfnamefont {A.}~\bibnamefont {Vaartjes}}, \bibinfo {author} {\bibfnamefont {S.}~\bibnamefont {Asaad}}, \bibinfo {author} {\bibfnamefont {V.}~\bibnamefont {Mourik}}, \bibinfo {author} {\bibfnamefont {F.~E.}\ \bibnamefont {Hudson}}, \bibinfo {author} {\bibfnamefont {K.~M.}\ \bibnamefont {Itoh}}, \bibinfo {author} {\bibfnamefont {B.~C.}\ \bibnamefont {Johnson}}, \bibinfo {author} {\bibfnamefont {A.~M.}\ \bibnamefont {Jakob}}, \bibinfo {author} {\bibfnamefont {J.~C.}\ \bibnamefont {McCallum}}, \bibinfo {author} {\bibfnamefont {D.~N.}\ \bibnamefont {Jamieson}}, \bibinfo {author} {\bibfnamefont {A.~S.}\ \bibnamefont {Dzurak}},\ and\ \bibinfo {author} {\bibfnamefont {A.}~\bibnamefont {Morello}},\ }\href {https://doi.org/10.1038/s41467-024-45368-y} {\bibfield
  {journal} {\bibinfo  {journal} {Nature Communications}\ }\textbf {\bibinfo {volume} {15}},\ \bibinfo {pages} {1380} (\bibinfo {year} {2024})},\ \bibinfo {note} {publisher: Nature Publishing Group}\BibitemShut {NoStop}%
\bibitem [{\citenamefont {Godfrin}\ \emph {et~al.}(2017)\citenamefont {Godfrin}, \citenamefont {Ferhat}, \citenamefont {Ballou}, \citenamefont {Klyatskaya}, \citenamefont {Ruben}, \citenamefont {Wernsdorfer},\ and\ \citenamefont {Balestro}}]{Godfrin2017}%
  \BibitemOpen
  \bibfield  {author} {\bibinfo {author} {\bibfnamefont {C.}~\bibnamefont {Godfrin}}, \bibinfo {author} {\bibfnamefont {A.}~\bibnamefont {Ferhat}}, \bibinfo {author} {\bibfnamefont {R.}~\bibnamefont {Ballou}}, \bibinfo {author} {\bibfnamefont {S.}~\bibnamefont {Klyatskaya}}, \bibinfo {author} {\bibfnamefont {M.}~\bibnamefont {Ruben}}, \bibinfo {author} {\bibfnamefont {W.}~\bibnamefont {Wernsdorfer}},\ and\ \bibinfo {author} {\bibfnamefont {F.}~\bibnamefont {Balestro}},\ }\href {https://doi.org/10.1103/PhysRevLett.119.187702} {\bibfield  {journal} {\bibinfo  {journal} {Phys. Rev. Lett.}\ }\textbf {\bibinfo {volume} {119}},\ \bibinfo {pages} {187702} (\bibinfo {year} {2017})}\BibitemShut {NoStop}%
\bibitem [{\citenamefont {Champion}\ \emph {et~al.}(2024)\citenamefont {Champion}, \citenamefont {Wang}, \citenamefont {Parker},\ and\ \citenamefont {Blok}}]{champion24}%
  \BibitemOpen
  \bibfield  {author} {\bibinfo {author} {\bibfnamefont {E.}~\bibnamefont {Champion}}, \bibinfo {author} {\bibfnamefont {Z.}~\bibnamefont {Wang}}, \bibinfo {author} {\bibfnamefont {R.}~\bibnamefont {Parker}},\ and\ \bibinfo {author} {\bibfnamefont {M.}~\bibnamefont {Blok}},\ }\href@noop {} {\bibfield  {journal} {\bibinfo  {journal} {arXiv preprint arXiv:2405.15857}\ } (\bibinfo {year} {2024})}\BibitemShut {NoStop}%
\bibitem [{\citenamefont {Goss}\ \emph {et~al.}(2022)\citenamefont {Goss}, \citenamefont {Morvan}, \citenamefont {Marinelli}, \citenamefont {Mitchell}, \citenamefont {Nguyen}, \citenamefont {Naik}, \citenamefont {Chen}, \citenamefont {J{\"u}nger}, \citenamefont {Kreikebaum}, \citenamefont {Santiago} \emph {et~al.}}]{Goss2022}%
  \BibitemOpen
  \bibfield  {author} {\bibinfo {author} {\bibfnamefont {N.}~\bibnamefont {Goss}}, \bibinfo {author} {\bibfnamefont {A.}~\bibnamefont {Morvan}}, \bibinfo {author} {\bibfnamefont {B.}~\bibnamefont {Marinelli}}, \bibinfo {author} {\bibfnamefont {B.~K.}\ \bibnamefont {Mitchell}}, \bibinfo {author} {\bibfnamefont {L.~B.}\ \bibnamefont {Nguyen}}, \bibinfo {author} {\bibfnamefont {R.~K.}\ \bibnamefont {Naik}}, \bibinfo {author} {\bibfnamefont {L.}~\bibnamefont {Chen}}, \bibinfo {author} {\bibfnamefont {C.}~\bibnamefont {J{\"u}nger}}, \bibinfo {author} {\bibfnamefont {J.~M.}\ \bibnamefont {Kreikebaum}}, \bibinfo {author} {\bibfnamefont {D.~I.}\ \bibnamefont {Santiago}}, \emph {et~al.},\ }\href@noop {} {\bibfield  {journal} {\bibinfo  {journal} {Nature communications}\ }\textbf {\bibinfo {volume} {13}},\ \bibinfo {pages} {7481} (\bibinfo {year} {2022})}\BibitemShut {NoStop}%
\bibitem [{\citenamefont {Fischer}\ \emph {et~al.}(2023)\citenamefont {Fischer}, \citenamefont {Chiesa}, \citenamefont {Tacchino}, \citenamefont {Egger}, \citenamefont {Carretta},\ and\ \citenamefont {Tavernelli}}]{Laurin2023}%
  \BibitemOpen
  \bibfield  {author} {\bibinfo {author} {\bibfnamefont {L.~E.}\ \bibnamefont {Fischer}}, \bibinfo {author} {\bibfnamefont {A.}~\bibnamefont {Chiesa}}, \bibinfo {author} {\bibfnamefont {F.}~\bibnamefont {Tacchino}}, \bibinfo {author} {\bibfnamefont {D.~J.}\ \bibnamefont {Egger}}, \bibinfo {author} {\bibfnamefont {S.}~\bibnamefont {Carretta}},\ and\ \bibinfo {author} {\bibfnamefont {I.}~\bibnamefont {Tavernelli}},\ }\href {https://doi.org/10.1103/PRXQuantum.4.030327} {\bibfield  {journal} {\bibinfo  {journal} {PRX Quantum}\ }\textbf {\bibinfo {volume} {4}},\ \bibinfo {pages} {030327} (\bibinfo {year} {2023})}\BibitemShut {NoStop}%
\bibitem [{\citenamefont {Neeley}\ \emph {et~al.}(2009)\citenamefont {Neeley}, \citenamefont {Ansmann}, \citenamefont {Bialczak}, \citenamefont {Hofheinz}, \citenamefont {Lucero}, \citenamefont {O'Connell}, \citenamefont {Sank}, \citenamefont {Wang}, \citenamefont {Wenner}, \citenamefont {Cleland} \emph {et~al.}}]{Neeley2009}%
  \BibitemOpen
  \bibfield  {author} {\bibinfo {author} {\bibfnamefont {M.}~\bibnamefont {Neeley}}, \bibinfo {author} {\bibfnamefont {M.}~\bibnamefont {Ansmann}}, \bibinfo {author} {\bibfnamefont {R.~C.}\ \bibnamefont {Bialczak}}, \bibinfo {author} {\bibfnamefont {M.}~\bibnamefont {Hofheinz}}, \bibinfo {author} {\bibfnamefont {E.}~\bibnamefont {Lucero}}, \bibinfo {author} {\bibfnamefont {A.~D.}\ \bibnamefont {O'Connell}}, \bibinfo {author} {\bibfnamefont {D.}~\bibnamefont {Sank}}, \bibinfo {author} {\bibfnamefont {H.}~\bibnamefont {Wang}}, \bibinfo {author} {\bibfnamefont {J.}~\bibnamefont {Wenner}}, \bibinfo {author} {\bibfnamefont {A.~N.}\ \bibnamefont {Cleland}}, \emph {et~al.},\ }\href@noop {} {\bibfield  {journal} {\bibinfo  {journal} {Science}\ }\textbf {\bibinfo {volume} {325}},\ \bibinfo {pages} {722} (\bibinfo {year} {2009})}\BibitemShut {NoStop}%
\bibitem [{\citenamefont {Chi}\ \emph {et~al.}(2022)\citenamefont {Chi}, \citenamefont {Huang}, \citenamefont {Zhang}, \citenamefont {Mao}, \citenamefont {Zhou}, \citenamefont {Chen}, \citenamefont {Zhai}, \citenamefont {Bao}, \citenamefont {Dai}, \citenamefont {Yuan} \emph {et~al.}}]{Chi2022}%
  \BibitemOpen
  \bibfield  {author} {\bibinfo {author} {\bibfnamefont {Y.}~\bibnamefont {Chi}}, \bibinfo {author} {\bibfnamefont {J.}~\bibnamefont {Huang}}, \bibinfo {author} {\bibfnamefont {Z.}~\bibnamefont {Zhang}}, \bibinfo {author} {\bibfnamefont {J.}~\bibnamefont {Mao}}, \bibinfo {author} {\bibfnamefont {Z.}~\bibnamefont {Zhou}}, \bibinfo {author} {\bibfnamefont {X.}~\bibnamefont {Chen}}, \bibinfo {author} {\bibfnamefont {C.}~\bibnamefont {Zhai}}, \bibinfo {author} {\bibfnamefont {J.}~\bibnamefont {Bao}}, \bibinfo {author} {\bibfnamefont {T.}~\bibnamefont {Dai}}, \bibinfo {author} {\bibfnamefont {H.}~\bibnamefont {Yuan}}, \emph {et~al.},\ }\href@noop {} {\bibfield  {journal} {\bibinfo  {journal} {Nature communications}\ }\textbf {\bibinfo {volume} {13}},\ \bibinfo {pages} {1166} (\bibinfo {year} {2022})}\BibitemShut {NoStop}%
\bibitem [{\citenamefont {Kues}\ \emph {et~al.}(2017)\citenamefont {Kues}, \citenamefont {Reimer}, \citenamefont {Roztocki}, \citenamefont {Cort{\'e}s}, \citenamefont {Sciara}, \citenamefont {Wetzel}, \citenamefont {Zhang}, \citenamefont {Cino}, \citenamefont {Chu}, \citenamefont {Little} \emph {et~al.}}]{Kues2017}%
  \BibitemOpen
  \bibfield  {author} {\bibinfo {author} {\bibfnamefont {M.}~\bibnamefont {Kues}}, \bibinfo {author} {\bibfnamefont {C.}~\bibnamefont {Reimer}}, \bibinfo {author} {\bibfnamefont {P.}~\bibnamefont {Roztocki}}, \bibinfo {author} {\bibfnamefont {L.~R.}\ \bibnamefont {Cort{\'e}s}}, \bibinfo {author} {\bibfnamefont {S.}~\bibnamefont {Sciara}}, \bibinfo {author} {\bibfnamefont {B.}~\bibnamefont {Wetzel}}, \bibinfo {author} {\bibfnamefont {Y.}~\bibnamefont {Zhang}}, \bibinfo {author} {\bibfnamefont {A.}~\bibnamefont {Cino}}, \bibinfo {author} {\bibfnamefont {S.~T.}\ \bibnamefont {Chu}}, \bibinfo {author} {\bibfnamefont {B.~E.}\ \bibnamefont {Little}}, \emph {et~al.},\ }\href@noop {} {\bibfield  {journal} {\bibinfo  {journal} {Nature}\ }\textbf {\bibinfo {volume} {546}},\ \bibinfo {pages} {622} (\bibinfo {year} {2017})}\BibitemShut {NoStop}%
\bibitem [{\citenamefont {Aksenov}\ \emph {et~al.}(2023)\citenamefont {Aksenov}, \citenamefont {Zalivako}, \citenamefont {Semerikov}, \citenamefont {Borisenko}, \citenamefont {Semenin}, \citenamefont {Sidorov}, \citenamefont {Fedorov}, \citenamefont {Khabarova},\ and\ \citenamefont {Kolachevsky}}]{Aksenov2023}%
  \BibitemOpen
  \bibfield  {author} {\bibinfo {author} {\bibfnamefont {M.~A.}\ \bibnamefont {Aksenov}}, \bibinfo {author} {\bibfnamefont {I.~V.}\ \bibnamefont {Zalivako}}, \bibinfo {author} {\bibfnamefont {I.~A.}\ \bibnamefont {Semerikov}}, \bibinfo {author} {\bibfnamefont {A.~S.}\ \bibnamefont {Borisenko}}, \bibinfo {author} {\bibfnamefont {N.~V.}\ \bibnamefont {Semenin}}, \bibinfo {author} {\bibfnamefont {P.~L.}\ \bibnamefont {Sidorov}}, \bibinfo {author} {\bibfnamefont {A.~K.}\ \bibnamefont {Fedorov}}, \bibinfo {author} {\bibfnamefont {K.~Y.}\ \bibnamefont {Khabarova}},\ and\ \bibinfo {author} {\bibfnamefont {N.~N.}\ \bibnamefont {Kolachevsky}},\ }\href {https://doi.org/10.1103/PhysRevA.107.052612} {\bibfield  {journal} {\bibinfo  {journal} {Phys. Rev. A}\ }\textbf {\bibinfo {volume} {107}},\ \bibinfo {pages} {052612} (\bibinfo {year} {2023})}\BibitemShut {NoStop}%
\bibitem [{\citenamefont {Hrmo}\ \emph {et~al.}(2023)\citenamefont {Hrmo}, \citenamefont {Wilhelm}, \citenamefont {Gerster}, \citenamefont {van Mourik}, \citenamefont {Huber}, \citenamefont {Blatt}, \citenamefont {Schindler}, \citenamefont {Monz},\ and\ \citenamefont {Ringbauer}}]{Hrmo2023}%
  \BibitemOpen
  \bibfield  {author} {\bibinfo {author} {\bibfnamefont {P.}~\bibnamefont {Hrmo}}, \bibinfo {author} {\bibfnamefont {B.}~\bibnamefont {Wilhelm}}, \bibinfo {author} {\bibfnamefont {L.}~\bibnamefont {Gerster}}, \bibinfo {author} {\bibfnamefont {M.~W.}\ \bibnamefont {van Mourik}}, \bibinfo {author} {\bibfnamefont {M.}~\bibnamefont {Huber}}, \bibinfo {author} {\bibfnamefont {R.}~\bibnamefont {Blatt}}, \bibinfo {author} {\bibfnamefont {P.}~\bibnamefont {Schindler}}, \bibinfo {author} {\bibfnamefont {T.}~\bibnamefont {Monz}},\ and\ \bibinfo {author} {\bibfnamefont {M.}~\bibnamefont {Ringbauer}},\ }\href@noop {} {\bibfield  {journal} {\bibinfo  {journal} {Nature Communications}\ }\textbf {\bibinfo {volume} {14}},\ \bibinfo {pages} {2242} (\bibinfo {year} {2023})}\BibitemShut {NoStop}%
\bibitem [{\citenamefont {Ringbauer}\ \emph {et~al.}(2022)\citenamefont {Ringbauer}, \citenamefont {Meth}, \citenamefont {Postler}, \citenamefont {Stricker}, \citenamefont {Blatt}, \citenamefont {Schindler},\ and\ \citenamefont {Monz}}]{Ringbauer2022}%
  \BibitemOpen
  \bibfield  {author} {\bibinfo {author} {\bibfnamefont {M.}~\bibnamefont {Ringbauer}}, \bibinfo {author} {\bibfnamefont {M.}~\bibnamefont {Meth}}, \bibinfo {author} {\bibfnamefont {L.}~\bibnamefont {Postler}}, \bibinfo {author} {\bibfnamefont {R.}~\bibnamefont {Stricker}}, \bibinfo {author} {\bibfnamefont {R.}~\bibnamefont {Blatt}}, \bibinfo {author} {\bibfnamefont {P.}~\bibnamefont {Schindler}},\ and\ \bibinfo {author} {\bibfnamefont {T.}~\bibnamefont {Monz}},\ }\href@noop {} {\bibfield  {journal} {\bibinfo  {journal} {Nature Physics}\ }\textbf {\bibinfo {volume} {18}},\ \bibinfo {pages} {1053} (\bibinfo {year} {2022})}\BibitemShut {NoStop}%
\bibitem [{\citenamefont {Brennen}\ \emph {et~al.}(2005)\citenamefont {Brennen}, \citenamefont {O'Leary},\ and\ \citenamefont {Bullock}}]{Gavin2005}%
  \BibitemOpen
  \bibfield  {author} {\bibinfo {author} {\bibfnamefont {G.~K.}\ \bibnamefont {Brennen}}, \bibinfo {author} {\bibfnamefont {D.~P.}\ \bibnamefont {O'Leary}},\ and\ \bibinfo {author} {\bibfnamefont {S.~S.}\ \bibnamefont {Bullock}},\ }\href {https://doi.org/10.1103/PhysRevA.71.052318} {\bibfield  {journal} {\bibinfo  {journal} {Phys. Rev. A}\ }\textbf {\bibinfo {volume} {71}},\ \bibinfo {pages} {052318} (\bibinfo {year} {2005})}\BibitemShut {NoStop}%
\bibitem [{\citenamefont {L{\"o}schnauer}\ \emph {et~al.}(2024)\citenamefont {L{\"o}schnauer}, \citenamefont {Toba}, \citenamefont {Hughes}, \citenamefont {King}, \citenamefont {Weber}, \citenamefont {Srinivas}, \citenamefont {Matt}, \citenamefont {Nourshargh}, \citenamefont {Allcock}, \citenamefont {Ballance} \emph {et~al.}}]{Loschnauer2024}%
  \BibitemOpen
  \bibfield  {author} {\bibinfo {author} {\bibfnamefont {C.}~\bibnamefont {L{\"o}schnauer}}, \bibinfo {author} {\bibfnamefont {J.~M.}\ \bibnamefont {Toba}}, \bibinfo {author} {\bibfnamefont {A.}~\bibnamefont {Hughes}}, \bibinfo {author} {\bibfnamefont {S.}~\bibnamefont {King}}, \bibinfo {author} {\bibfnamefont {M.}~\bibnamefont {Weber}}, \bibinfo {author} {\bibfnamefont {R.}~\bibnamefont {Srinivas}}, \bibinfo {author} {\bibfnamefont {R.}~\bibnamefont {Matt}}, \bibinfo {author} {\bibfnamefont {R.}~\bibnamefont {Nourshargh}}, \bibinfo {author} {\bibfnamefont {D.}~\bibnamefont {Allcock}}, \bibinfo {author} {\bibfnamefont {C.}~\bibnamefont {Ballance}}, \emph {et~al.},\ }\href@noop {} {\bibfield  {journal} {\bibinfo  {journal} {arXiv preprint arXiv:2407.07694}\ } (\bibinfo {year} {2024})}\BibitemShut {NoStop}%
\bibitem [{\citenamefont {Ballance}\ \emph {et~al.}(2016)\citenamefont {Ballance}, \citenamefont {Harty}, \citenamefont {Linke}, \citenamefont {Sepiol},\ and\ \citenamefont {Lucas}}]{Ballance2016}%
  \BibitemOpen
  \bibfield  {author} {\bibinfo {author} {\bibfnamefont {C.~J.}\ \bibnamefont {Ballance}}, \bibinfo {author} {\bibfnamefont {T.~P.}\ \bibnamefont {Harty}}, \bibinfo {author} {\bibfnamefont {N.~M.}\ \bibnamefont {Linke}}, \bibinfo {author} {\bibfnamefont {M.~A.}\ \bibnamefont {Sepiol}},\ and\ \bibinfo {author} {\bibfnamefont {D.~M.}\ \bibnamefont {Lucas}},\ }\href {https://doi.org/10.1103/PhysRevLett.117.060504} {\bibfield  {journal} {\bibinfo  {journal} {Phys. Rev. Lett.}\ }\textbf {\bibinfo {volume} {117}},\ \bibinfo {pages} {060504} (\bibinfo {year} {2016})}\BibitemShut {NoStop}%
\bibitem [{\citenamefont {Low}\ \emph {et~al.}(2025)\citenamefont {Low}, \citenamefont {White},\ and\ \citenamefont {Senko}}]{Low2025}%
  \BibitemOpen
  \bibfield  {author} {\bibinfo {author} {\bibfnamefont {P.~J.}\ \bibnamefont {Low}}, \bibinfo {author} {\bibfnamefont {B.}~\bibnamefont {White}},\ and\ \bibinfo {author} {\bibfnamefont {C.}~\bibnamefont {Senko}},\ }\href@noop {} {\bibfield  {journal} {\bibinfo  {journal} {npj Quantum Information}\ }\textbf {\bibinfo {volume} {11}},\ \bibinfo {pages} {1} (\bibinfo {year} {2025})}\BibitemShut {NoStop}%
\bibitem [{sup()}]{supp_material}%
  \BibitemOpen
  \href@noop {} {\bibinfo {title} {See supplementary material at [url] for randomized benchmarking, coherence time measurments, and pulse parameters for the \text{G}rover's search algorithm.}}\BibitemShut {Stop}%
\bibitem [{\citenamefont {Shi}\ \emph {et~al.}(2025)\citenamefont {Shi}, \citenamefont {Sinanan-Singh}, \citenamefont {DeBry}, \citenamefont {Todaro}, \citenamefont {Chuang},\ and\ \citenamefont {Chiaverini}}]{Shi2025}%
  \BibitemOpen
  \bibfield  {author} {\bibinfo {author} {\bibfnamefont {X.}~\bibnamefont {Shi}}, \bibinfo {author} {\bibfnamefont {J.}~\bibnamefont {Sinanan-Singh}}, \bibinfo {author} {\bibfnamefont {K.}~\bibnamefont {DeBry}}, \bibinfo {author} {\bibfnamefont {S.~L.}\ \bibnamefont {Todaro}}, \bibinfo {author} {\bibfnamefont {I.~L.}\ \bibnamefont {Chuang}},\ and\ \bibinfo {author} {\bibfnamefont {J.}~\bibnamefont {Chiaverini}},\ }\href {https://doi.org/10.1103/PhysRevA.111.L020601} {\bibfield  {journal} {\bibinfo  {journal} {Phys. Rev. A}\ }\textbf {\bibinfo {volume} {111}},\ \bibinfo {pages} {L020601} (\bibinfo {year} {2025})}\BibitemShut {NoStop}%
\bibitem [{\citenamefont {An}\ \emph {et~al.}(2022)\citenamefont {An}, \citenamefont {Ransford}, \citenamefont {Schaffer}, \citenamefont {Sletten}, \citenamefont {Gaebler}, \citenamefont {Hostetter},\ and\ \citenamefont {Vittorini}}]{An2022}%
  \BibitemOpen
  \bibfield  {author} {\bibinfo {author} {\bibfnamefont {F.~A.}\ \bibnamefont {An}}, \bibinfo {author} {\bibfnamefont {A.}~\bibnamefont {Ransford}}, \bibinfo {author} {\bibfnamefont {A.}~\bibnamefont {Schaffer}}, \bibinfo {author} {\bibfnamefont {L.~R.}\ \bibnamefont {Sletten}}, \bibinfo {author} {\bibfnamefont {J.}~\bibnamefont {Gaebler}}, \bibinfo {author} {\bibfnamefont {J.}~\bibnamefont {Hostetter}},\ and\ \bibinfo {author} {\bibfnamefont {G.}~\bibnamefont {Vittorini}},\ }\href {https://doi.org/10.1103/PhysRevLett.129.130501} {\bibfield  {journal} {\bibinfo  {journal} {Phys. Rev. Lett.}\ }\textbf {\bibinfo {volume} {129}},\ \bibinfo {pages} {130501} (\bibinfo {year} {2022})}\BibitemShut {NoStop}%
\bibitem [{\citenamefont {Sotirova}\ \emph {et~al.}(2024)\citenamefont {Sotirova}, \citenamefont {Leppard}, \citenamefont {Vazquez-Brennan}, \citenamefont {Decoppet}, \citenamefont {Pokorny}, \citenamefont {Malinowski},\ and\ \citenamefont {Ballance}}]{Sotirova2024}%
  \BibitemOpen
  \bibfield  {author} {\bibinfo {author} {\bibfnamefont {A.}~\bibnamefont {Sotirova}}, \bibinfo {author} {\bibfnamefont {J.}~\bibnamefont {Leppard}}, \bibinfo {author} {\bibfnamefont {A.}~\bibnamefont {Vazquez-Brennan}}, \bibinfo {author} {\bibfnamefont {S.}~\bibnamefont {Decoppet}}, \bibinfo {author} {\bibfnamefont {F.}~\bibnamefont {Pokorny}}, \bibinfo {author} {\bibfnamefont {M.}~\bibnamefont {Malinowski}},\ and\ \bibinfo {author} {\bibfnamefont {C.}~\bibnamefont {Ballance}},\ }\href@noop {} {\bibfield  {journal} {\bibinfo  {journal} {arXiv preprint arXiv:2409.05805}\ } (\bibinfo {year} {2024})}\BibitemShut {NoStop}%
\bibitem [{\citenamefont {Hradil}(1997)}]{Hradi1997}%
  \BibitemOpen
  \bibfield  {author} {\bibinfo {author} {\bibfnamefont {Z.}~\bibnamefont {Hradil}},\ }\href {https://doi.org/10.1103/PhysRevA.55.R1561} {\bibfield  {journal} {\bibinfo  {journal} {Phys. Rev. A}\ }\textbf {\bibinfo {volume} {55}},\ \bibinfo {pages} {R1561} (\bibinfo {year} {1997})}\BibitemShut {NoStop}%
\bibitem [{\citenamefont {Roy}\ \emph {et~al.}(2020)\citenamefont {Roy}, \citenamefont {Hazra}, \citenamefont {Kundu}, \citenamefont {Chand}, \citenamefont {Patankar},\ and\ \citenamefont {Vijay}}]{Tanay2020}%
  \BibitemOpen
  \bibfield  {author} {\bibinfo {author} {\bibfnamefont {T.}~\bibnamefont {Roy}}, \bibinfo {author} {\bibfnamefont {S.}~\bibnamefont {Hazra}}, \bibinfo {author} {\bibfnamefont {S.}~\bibnamefont {Kundu}}, \bibinfo {author} {\bibfnamefont {M.}~\bibnamefont {Chand}}, \bibinfo {author} {\bibfnamefont {M.~P.}\ \bibnamefont {Patankar}},\ and\ \bibinfo {author} {\bibfnamefont {R.}~\bibnamefont {Vijay}},\ }\href {https://doi.org/10.1103/PhysRevApplied.14.014072} {\bibfield  {journal} {\bibinfo  {journal} {Phys. Rev. Appl.}\ }\textbf {\bibinfo {volume} {14}},\ \bibinfo {pages} {014072} (\bibinfo {year} {2020})}\BibitemShut {NoStop}%
\bibitem [{\citenamefont {Leuenberger}\ and\ \citenamefont {Loss}(2003)}]{leuenberger03}%
  \BibitemOpen
  \bibfield  {author} {\bibinfo {author} {\bibfnamefont {M.~N.}\ \bibnamefont {Leuenberger}}\ and\ \bibinfo {author} {\bibfnamefont {D.}~\bibnamefont {Loss}},\ }\href {https://doi.org/10.1103/PhysRevB.68.165317} {\bibfield  {journal} {\bibinfo  {journal} {Physical Review B}\ }\textbf {\bibinfo {volume} {68}},\ \bibinfo {pages} {165317} (\bibinfo {year} {2003})},\ \bibinfo {note} {arXiv:cond-mat/0304674}\BibitemShut {NoStop}%
\bibitem [{\citenamefont {McKay}\ \emph {et~al.}(2017)\citenamefont {McKay}, \citenamefont {Wood}, \citenamefont {Sheldon}, \citenamefont {Chow},\ and\ \citenamefont {Gambetta}}]{mckay_efficient_2017}%
  \BibitemOpen
  \bibfield  {author} {\bibinfo {author} {\bibfnamefont {D.~C.}\ \bibnamefont {McKay}}, \bibinfo {author} {\bibfnamefont {C.~J.}\ \bibnamefont {Wood}}, \bibinfo {author} {\bibfnamefont {S.}~\bibnamefont {Sheldon}}, \bibinfo {author} {\bibfnamefont {J.~M.}\ \bibnamefont {Chow}},\ and\ \bibinfo {author} {\bibfnamefont {J.~M.}\ \bibnamefont {Gambetta}},\ }\href {https://doi.org/10.1103/PhysRevA.96.022330} {\bibfield  {journal} {\bibinfo  {journal} {Physical Review A}\ }\textbf {\bibinfo {volume} {96}},\ \bibinfo {pages} {022330} (\bibinfo {year} {2017})}\BibitemShut {NoStop}%
\bibitem [{\citenamefont {Bullock}\ \emph {et~al.}(2005)\citenamefont {Bullock}, \citenamefont {O’Leary},\ and\ \citenamefont {Brennen}}]{Bullock2005}%
  \BibitemOpen
  \bibfield  {author} {\bibinfo {author} {\bibfnamefont {S.~S.}\ \bibnamefont {Bullock}}, \bibinfo {author} {\bibfnamefont {D.~P.}\ \bibnamefont {O’Leary}},\ and\ \bibinfo {author} {\bibfnamefont {G.~K.}\ \bibnamefont {Brennen}},\ }\href@noop {} {\bibfield  {journal} {\bibinfo  {journal} {Physical review letters}\ }\textbf {\bibinfo {volume} {94}},\ \bibinfo {pages} {230502} (\bibinfo {year} {2005})}\BibitemShut {NoStop}%
\end{thebibliography}%

\clearpage
\onecolumngrid
\section{Supplementary Material}

\subsection{Randomized Benchmarking}
To perform randomized benchmarking (RB), we randomly pick a sequence of gates from the qubit Clifford group embedded in SU$(d)$ as the irreducible weight $d$ representation of SU$(2)$. The 24 elements in this group are generated by the Pauli (I, X, Y, Z), Hadamard (H), and Phase (S) gates, and each element is decomposed into $1-2$ native spin-displacement $\pi$-pulses. 

\begin{figure}[htp]
    \centering
    \includegraphics[width = 0.5\textwidth]{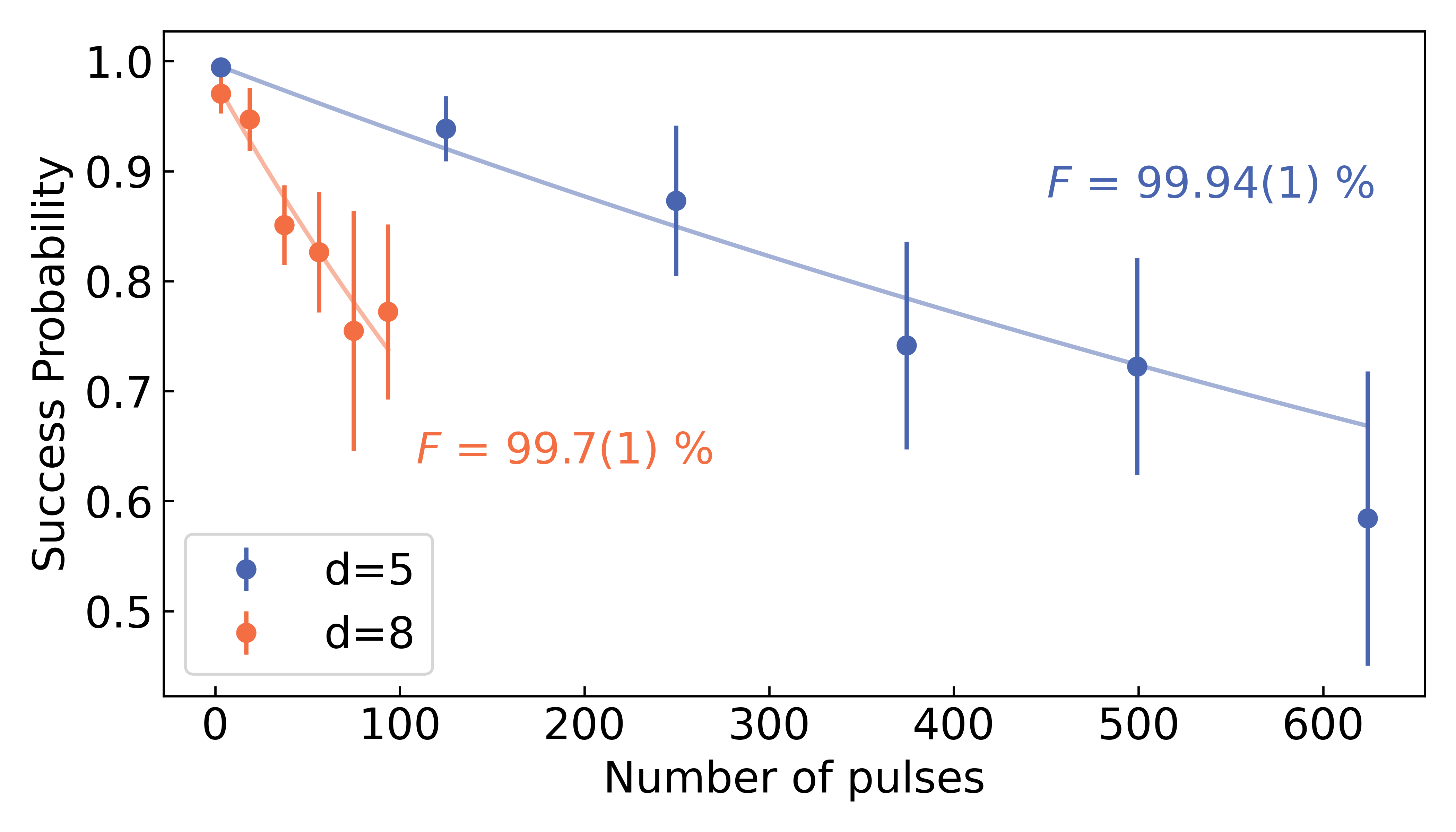}
    \caption{Randomized benchmarking of SU(2) operations on the qudit. We fit the data to exponential decay and extract the pulse fidelity to be 99.94(1)$\%$ and 99.7(1)$\%$ for the $d = 5$ and $d = 8$ cases, respectively. The error bars represent 68$\%$ confidence level.}
    \label{fig:rb}
\end{figure}

The benchmarking result is shown in Fig.~\ref{fig:rb}. According to Grover's search experiments, the measured pulse fidelity is approximately three times lower than the error rate we'd expect. This could be due to the fact that we are not performing SU(2) gates for the algorithm operations, and suggests that a better benchmarking method should be used to characterize the error rate in this case.

\subsection{Calibration}

\begin{figure}[htp]
    \centering

    \begin{minipage}[t]{0.49\textwidth}
        \centering
        (a) Individual RB Landscapes
        \vspace{4pt}

        \begin{minipage}[t]{0.48\textwidth}
            \includegraphics[width=\linewidth]{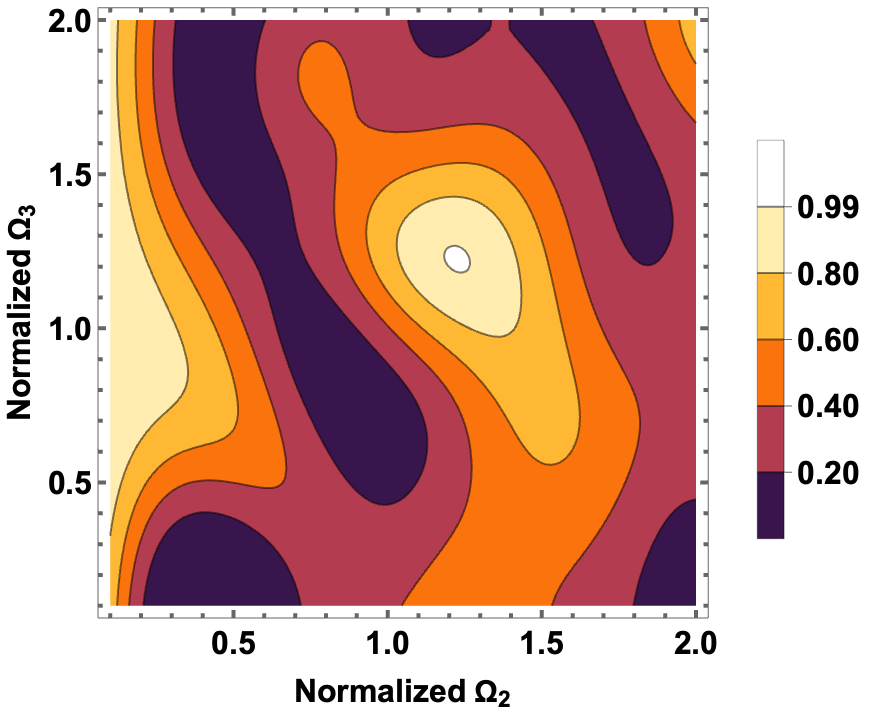} \\
            \vspace{2pt}
            \includegraphics[width=\linewidth]{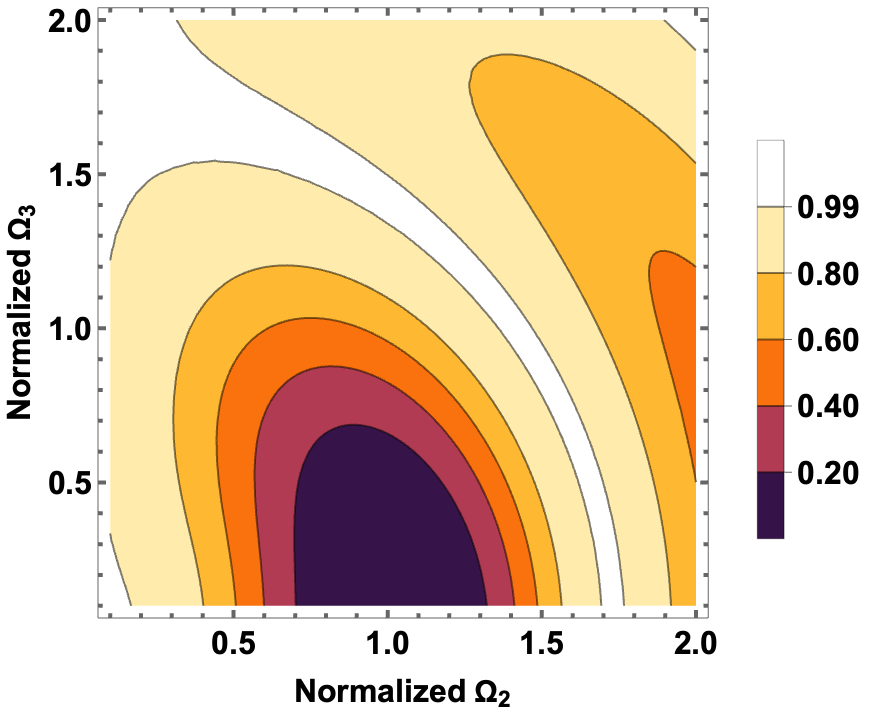}
        \end{minipage}
        \hfill
        \begin{minipage}[t]{0.48\textwidth}
            \includegraphics[width=\linewidth]{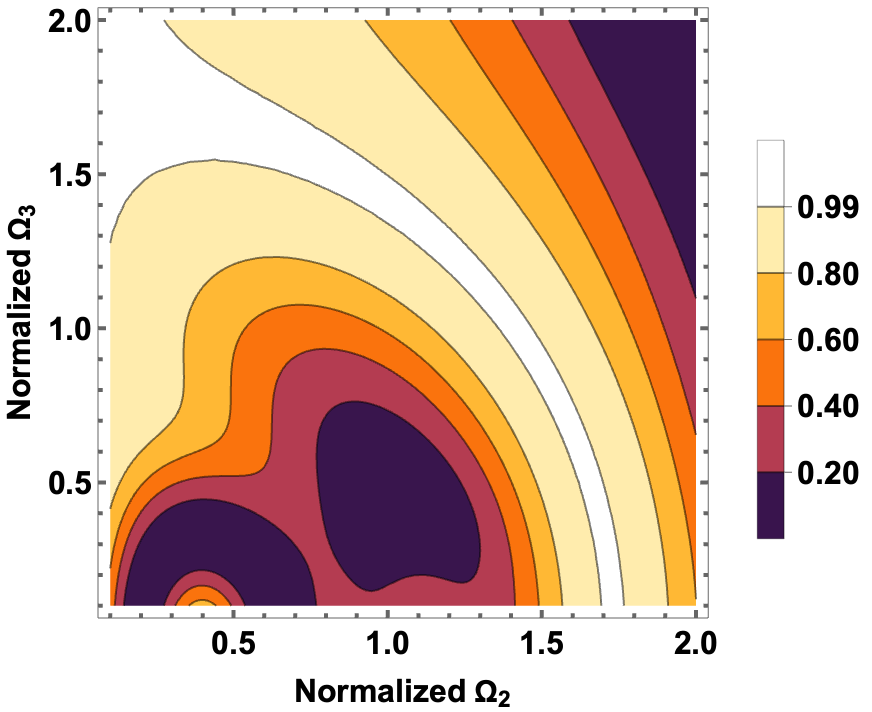} \\
            \vspace{2pt}
            \includegraphics[width=\linewidth]{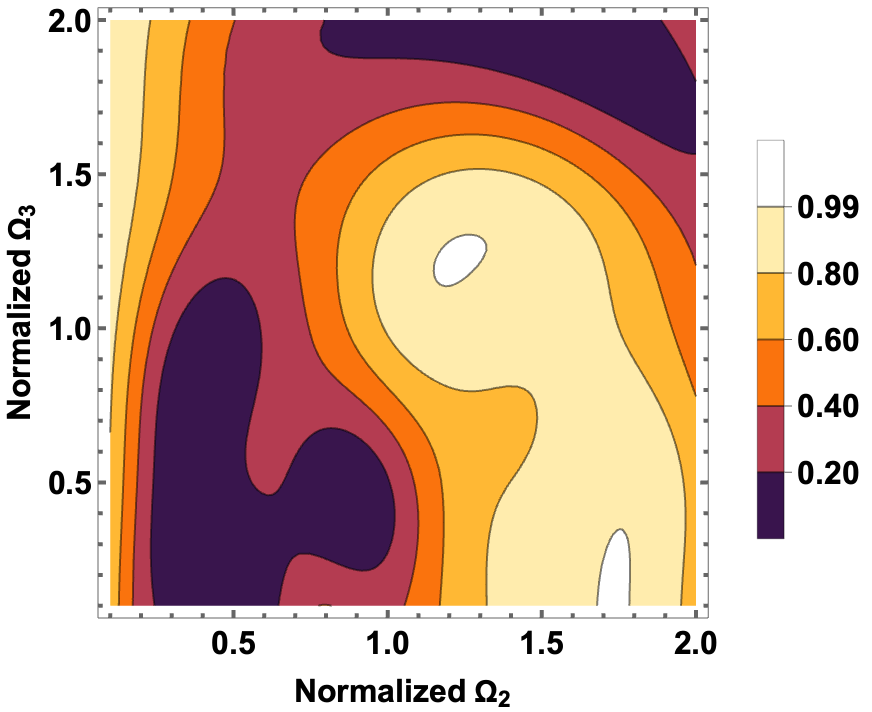}
        \end{minipage}
    \end{minipage}
    \hfill
    \begin{minipage}[t]{0.49\textwidth}
        \centering
        (b) Averaged RB Landscape
        \vspace{4pt}

        \includegraphics[width=\linewidth]{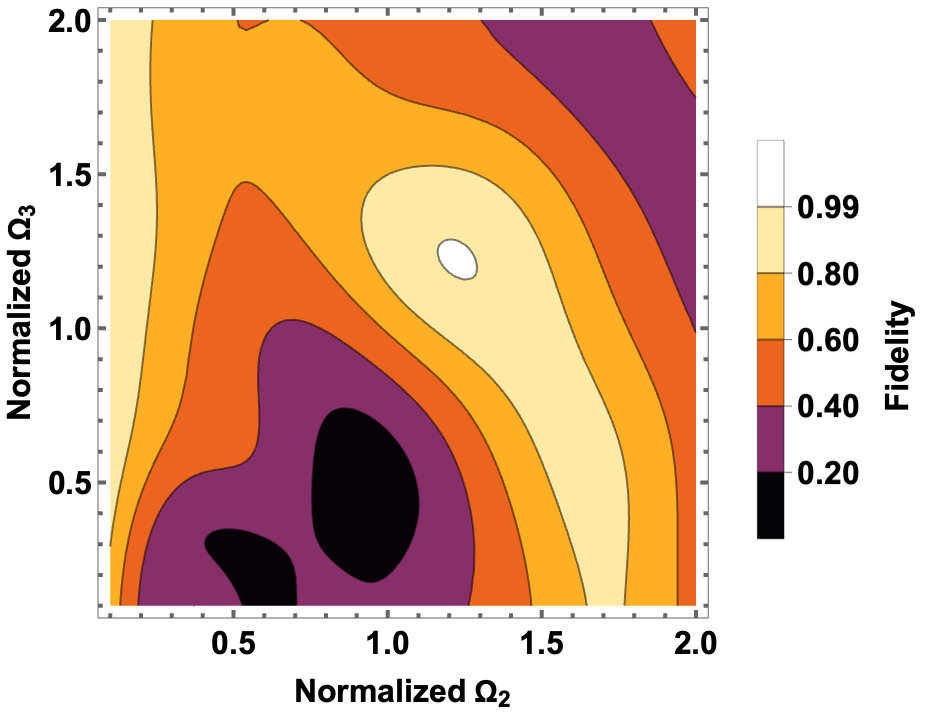}
    \end{minipage}

    \caption{Individual randomized benchmarking (RB) landscapes. (a) Fidelity as the RF amplitude (expressed as normalized Rabi frequencies) of the second and third tones are varied while the amplitudes of first and last tones are fixed at the optimal values for a spin-displacement pulse. In (b), the averaged landscape from these four sequences is shown.}
    \label{fig:rb_landscapes}
\end{figure}

We average many random sequences of global SU(2) Clifford gates (benchmarking sequences) to calibrate the displacement $\pi$-pulse to the Bloch equator in order to avoid minima valleys in the optimization landscape. Let's consider an example in d$=5$ where we are calibrating the RF amplitudes of four tones. For the purpose of visualization, suppose two tones begin well-calibrated, and the other two tones are being optimized via the RB. In Fig.~\ref{fig:rb_landscapes}, contour plots of RB sequences' fidelity with respect to the Rabi amplitudes of two tones show that valleys of minima are possible in individual sequences. However, averaging fixes this issue and creates a global minimum in the tuning range.

In general, all tones need to be calibrated simultaneously as the parameters are interdependent and the landscape highly non-linear. Averaging the fidelity of a few random RB sequences of different lengths makes both the calibration process and estimation of average gate error more accurate.

\subsection{Coherence time}
\label{app:coherence}
To measure the coherence time, a Ramsey-type experiment for the qudit, where the $\frac{\pi}{2}$ pulses are now displacement $\frac{\pi}{2}$ pulses, could be performed to extract a coherent time. After the sequence, we measure the population of each state and take the average weighted by their assigned $J_z$ value of $-\frac{d}{2}+i$ as the expectation value of $J_z$. The decaying oscillation amplitude is fitted to master equation simulation, with the same dephasing operator mentioned in the main text. The input parameters are the detuning of each tone and the coherence time. The shorter coherence time is expected due to the higher magnetic field sensitivities for the d = 8 qudit. The sum of the absolute value of the sensitivity is $\mathrm{MHz}/\mathrm{G}$, while it's $\mathrm{MHz}/\mathrm{G}$ for the d = 5 case. It's possible to find a set of states with lower magnetic field sensitivities at the cost of other properties of the qudit, such as transition frequency separation or transition strength.

\begin{figure}[htp]
    \centering
    \includegraphics[width = 0.5\textwidth]{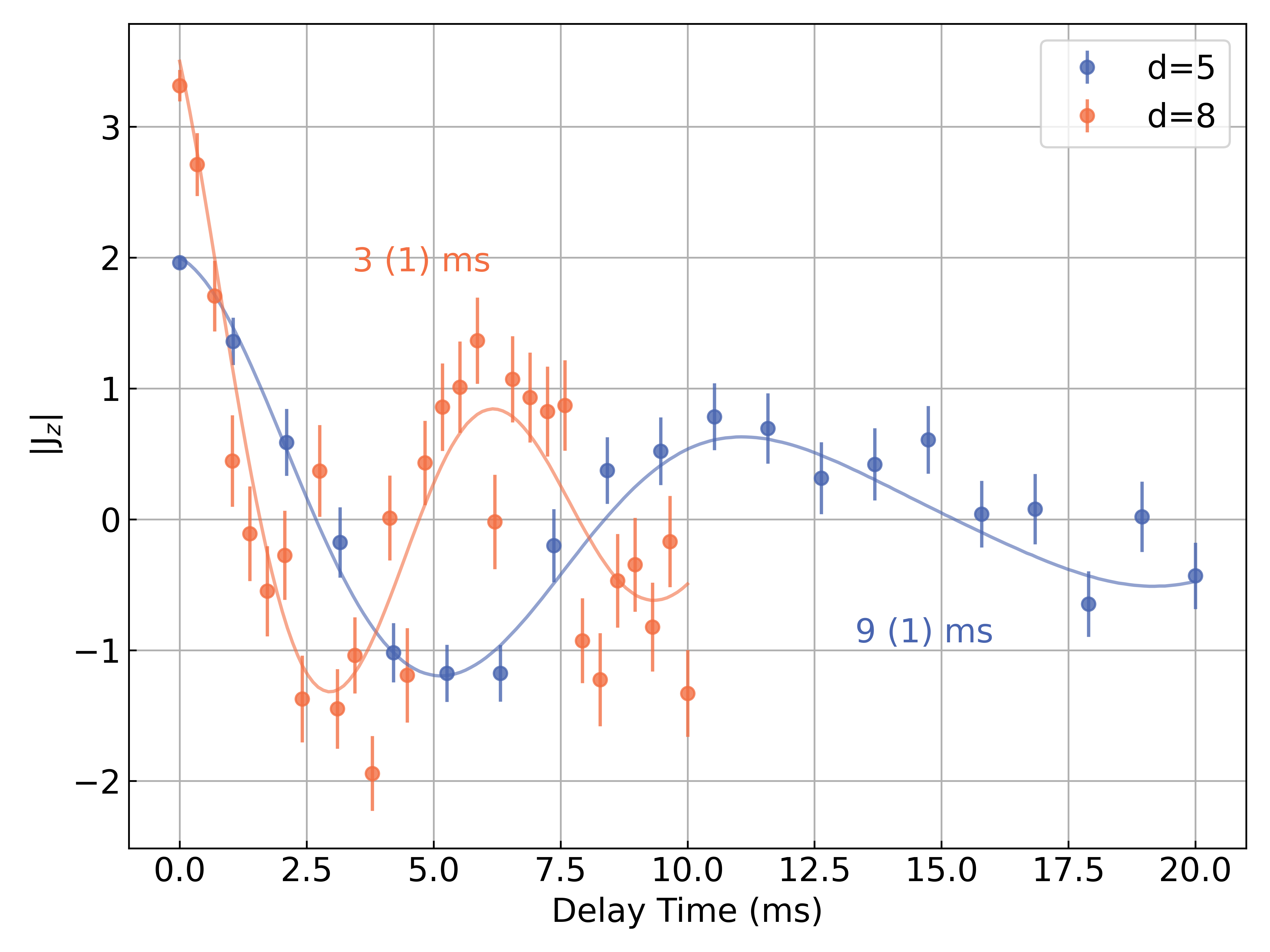}
    \caption{Ramsey-type experiment for the qudit. We extract a coherence time of 12(2) and 4.9(5) ms for the d = 5 and d = 8 qudits by fitting an exponential decay to the amplitude of oscillation of the artificial $J_z$.}
    \label{fig:ramsey}
\end{figure}

\subsection{Algorithm Sequence Parameters}

\label{app:grover_parameters}

In this section, we list the parameters (in order of phase oracle, preparation of equal superposition, and reflection) of the pulses used to implement Grover's search algorithm for a d = 5 and d = 8 qudit. The parameters include the displacement angle ($\theta = \Omega t$) and the phases of each tone, in units of radians.

    \begin{table}[h]
    \centering
    \caption{Pulse parameters for d=5}
    \label{tab:quantum_pulses}
    \begin{tabular}{ccS[table-format=1.4]S[table-format=-1.4]S[table-format=-1.4]S[table-format=-1.4]S[table-format=-1.4]}
    \toprule
    \textbf{Operation} & \textbf{Pulse} & \textbf{$\theta$} & \textbf{$\varphi_{1}$} & \textbf{$\varphi_{2}$} & \textbf{$\varphi_{3}$} & \textbf{$\varphi_{4}$} \\
    \midrule
    
    \multirow{2}{*}{Mark 0} & 1 & 1.8486 & 1.8390 & 1.9759 & -1.7224 & 1.3997 \\
     & 2 & 1.8476 & 1.8403 & -1.1644 & 1.4201 & -1.7405 \\
     \midrule
    \multirow{2}{*}{Mark 1}
    & 1 & 0.3995 & 1.7577 & -1.1708 & 0.0936 & -0.2708 \\
    & 2 & 0.4003 & 1.7574 & 5.1143 & 3.2356 & 2.8702 \\
     \midrule
    \multirow{2}{*}{Mark 2} & 1 & 0.4206 & 3.1942 & 2.2381 & -0.8326 & -0.8528 \\
     & 2 & 0.4201 & 0.0538 & 2.2378 & -0.8329 & -3.9939 \\
     \midrule
    \multirow{2}{*}{Mark 3} & 1 & 2.4190 & -0.9448 & -3.4755 & 1.3757 & -3.0336 \\
     & 2 & 2.4163 & 2.1940 & -0.3374 & 1.3727 & -3.0350 \\
     \midrule
    \multirow{2}{*}{Mark 4} 
    & 1 & 1.9584 & 1.1932 & 1.5179 & 2.2601 & 3.8294 \\
    & 2 & 1.9546 & -1.9522 & 4.6558 & 5.3980 & 3.8253 \\
    \midrule
    \multirow{2}{*}{Equal Sup.} 
    & 1 & 1.4919 & 0.7079 & 4.8253 & 5.3554 & 4.6164 \\
    & 2 & 1.8489 & -0.6297 & -0.1172 & 1.3818 & -1.0868 \\
    \midrule
    \multirow{4}{*}{Reflection} 
    & 1 & 1.2499 & 2.0320 & 0.3337 & 3.4753 & -1.1096 \\
    & 2 & 1.7134 & 1.4918 & -3.8264 & -0.6848 & 4.6334 \\
    & 3 & 0.8378 & -0.1852 & 3.0199 & 6.1615 & 2.9564 \\
    & 4 & 0.8296 & -0.5001 & -1.1443 & 1.9973 & 2.6415 \\
    \bottomrule
    \end{tabular}
    \end{table}

    \begin{table*}[htbp!]
    \centering
    \caption{Pulse parameters for d=8}
    \label{tab:quantum_pulses_7phase}
   \begin{tabular}{ccS[table-format=1.4]S[table-format=-1.4]S[table-format=-1.4]S[table-format=-1.4]S[table-format=-1.4]S[table-format=-1.4]S[table-format=-1.4]S[table-format=-1.4]}
    \toprule
    \textbf{Operation} & \textbf{Pulse} & \textbf{$\theta$} & \textbf{$\varphi_{1}$} & \textbf{$\varphi_{2}$} & \textbf{$\varphi_{3}$} & \textbf{$\varphi_{4}$} & \textbf{$\varphi_{5}$} & \textbf{$\varphi_{6}$} & \textbf{$\varphi_{7}$} \\
    \midrule
    \multirow{2}{*}{Mark 0} 
    & 1 & 1.6870 & 2.8899 & 1.8528 & -2.2374 & -0.4809 & -0.8473 & 0.5904 & 0.8701 \\
    & 2 & 1.6896 & 2.8938 & 4.9961 & 0.9040 & 2.6616 & 2.2942 & 3.7323 & 4.0122 \\
    \midrule
    \multirow{2}{*}{Mark 1} 
     & 1 & 0.4891 & 2.2565 & -0.0942 & -1.5233 & -2.2268 & -1.7298 & -3.4344 & 1.6510 \\
     & 2 & 0.4901 & 2.2562 & -0.0940 & 1.6158 & 0.9159 & 1.4116 & -0.2951 & -1.4909 \\
    \midrule
    \multirow{2}{*}{Mark 2} 
     & 1 & 0.2626 & 1.5553 & -4.6646 & -1.3044 & -0.2466 & -2.0645 & 1.3321 & 2.0275 \\
     & 2 & 0.2611 & -1.5882 & 1.6204 & 4.9801 & 2.8951 & 1.0772 & -1.8105 & -1.1152 \\
    \midrule
    \multirow{2}{*}{Mark 3} 
     & 1 & 0.2421 & 3.1602 & -0.7810 & 0.7803 & -2.1033 & 0.8550 & -0.9876 & 1.9103 \\
     & 2 & 0.2433 & 0.0175 & -3.9237 & 0.7800 & -2.1025 & -2.2852 & 2.1528 & -1.2323 \\
    \midrule
    \multirow{2}{*}{Mark 4} 
     & 1 & 0.8924 & 0.2175 & -1.8727 & 2.5708 & -0.4749 & 0.4002 & -4.0446 & -0.2344 \\
     & 2 & 1.1646 & -1.2161 & -0.1248 & -4.9105 & 0.9006 & 4.2297 & -0.4628 & -2.9096 \\
    \midrule
    \multirow{2}{*}{Mark 5} 
     & 1 & 0.2970 & 1.5332 & 1.3587 & -3.6896 & 0.6724 & 0.4276 & 4.1205 & 1.2141 \\
     & 2 & 0.2981 & -1.6075 & -1.7811 & -0.5479 & -2.4671 & 0.4274 & -2.1617 & -1.9276 \\
    \midrule
    \multirow{2}{*}{Mark 6} 
    & 1 & 0.2725 & 6.0795 & 1.2755 & 4.7039 & 4.4657 & 4.4599 & 1.9508 & 5.2308 \\
    & 2 & 0.2725 & 2.9366 & 4.4164 & 1.5618 & 1.3220 & 1.3179 & 1.9496 & -1.0504 \\
    \midrule
    \multirow{2}{*}{Mark 7} 
     & 1 & 1.5708 & 0 & 0 & 0 & 0 & 0 & 0 & 0 \\
     & 2 & 1.5708 & 0 & 0 & 0 & 0 & 0 & 0 & -3.1416 \\
    
    \midrule
    \multirow{3}{*}{Equal Sup.}
    & 1 & 1.0581 & 1.4234 & 1.8931 & 4.5766 & -1.4745 & -1.5473 & -1.5900 & -1.8468 \\
    & 2 & 0.7715 & -1.1582 & 2.9109 & -1.5840 & -0.9257 & -1.1627 & -1.8208 & -1.8958 \\
    & 3 & 0.3590 & -1.4932 & -1.3047 & 2.3407 & 2.6508 & 3.1032 & 3.3353 & -1.9226 \\
    
     \midrule
    
    \multirow{8}{*}{Reflection} 
    & 1 & 2.0303 & 1.1050 & 4.6208 & 4.9003 & 3.9970 & -0.6779 & 1.1179 & 0.6350 \\
    & 2 & 0.3594 & 2.5851 & 5.0821 & 5.9183 & 4.8872 & 4.6204 & 0.4102 & 3.5142 \\
    & 3 & 0.3393 & 0.9600 & 0.3362 & 6.2058 & 5.9570 & 2.6139 & 4.5210 & 1.4309 \\
    & 4 & 0.6325 & 2.4723 & 2.0203 & 6.7822 & 2.5239 & 0.6912 & -0.2383 & 4.5633 \\
    & 5 & 2.0861 & 7.3936 & -0.1733 & 1.1062 & 5.4852 & -1.8695 & 2.9519 & 3.6930 \\
    & 6 & 1.3068 & 0.3658 & 5.0804 & 4.7909 & 3.3969 & 7.0912 & -0.1226 & 4.0258 \\
    & 7 & 0.9287 & 0.2907 & 0.8797 & -0.8273 & 0.2935 & -2.9503 & -3.0324 & 1.8483 \\
    & 8 & 1.6214 & 3.4401 & 6.4809 & 3.7060 & 5.1973 & 3.3342 & 4.2009 & 4.5210 \\
    \bottomrule
    \end{tabular}
    \end{table*}

\FloatBarrier 

\end{document}